\documentclass[10pt]{article}
\usepackage[font=small]{caption}
\usepackage[utf8]{inputenc}
\usepackage{bbold}
\usepackage[a4paper, hmargin=2cm, tmargin=1.5cm, bmargin=3cm, footskip=2cm]{geometry}
\usepackage{mathtools}
\usepackage{float}
\usepackage{changepage}
\usepackage{dsfont}
\usepackage{secdot}
\usepackage[T1]{fontenc}
\usepackage{indentfirst}
\usepackage[inline]{enumitem}
\usepackage{subcaption}
\usepackage{amssymb}
\usepackage{amsfonts}
\usepackage{amsmath}
\usepackage{bm}
\usepackage{physics}
\usepackage{tensor}
\usepackage{hyperref}
\usepackage{color,soul}
\usepackage{siunitx}
\usepackage{mathrsfs}
\usepackage{cite}
\usepackage{authblk}

\AtBeginDocument{\RenewCommandCopy\qty\SI}

\DeclarePairedDelimiterX\sbraket[2]{[}{]}{#1\,\delimsize\vert\,\mathopen{}#2}
\DeclarePairedDelimiterX\sabraket[2]{[}{\rangle}{#1\,\delimsize\vert\,\mathopen{}#2}
\DeclarePairedDelimiterX\asbraket[2]{\langle}{]}{#1\,\delimsize\vert\,\mathopen{}#2}
\DeclarePairedDelimiterX\smel[3]{[}{]}{#1\,\delimsize\vert\,\mathopen{}#2\,\delimsize\vert\,\mathopen{}#3}
\DeclarePairedDelimiterX\samel[3]{[}{\rangle}{#1\,\delimsize\vert\,\mathopen{}#2\,\delimsize\vert\,\mathopen{}#3}
\DeclarePairedDelimiterX\asmel[3]{\langle}{]}{#1\,\delimsize\vert\,\mathopen{}#2\,\delimsize\vert\,\mathopen{}#3}

\newcommand{\id}{\mathbb{1}}
\newcommand{\lr}[1]{\left(#1\right)}
\newcommand{\myvec}[1]{\ensuremath{\begin{pmatrix}#1\end{pmatrix}}}

\begin{document}
\thispagestyle{empty}
\counterwithin{equation}{section}

\title{Probing Gluon TMD Models with Drell--Yan Structure Functions}

\author[1]{Jan Ferdyan}

\affil[1]{Institute of Theoretical Physics, Jagiellonian University, \protect\\
prof. {\L}ojasiewicza 11, 30-348 Kraków, Poland, \protect\\
email: jan.ferdyan@doctoral.uj.edu.pl}

\date{\today}
\maketitle

\counterwithin{equation}{section}

\abstract{
We compute structure functions for the Drell--Yan process in proton-proton collisions at the center of mass energy $\sqrt{S} = \qty{8}{\TeV}$, both parity conserving and parity breaking. For this calculation, we use the high-energy factorization formalism. The hard scattering matrix elements used in our derivation consist of two channels --- $q_\mathrm{val} g^* \to q V^*$ and $g^* g^* \to q \overline{q} V^*$, both at the tree level. We consider four types of gluon TMD models: Gaussian, Weizs\"{a}cker--Williams (WW), Kimber--Martin--Ryskin (KMR), and Jung--Hautmann (JH). We also consider the models with phenomenological adjustments to improve the data description. We derive and compare the structure functions calculated for different gluon TMD models with the ATLAS 2016 data. Based on this comparison, we calculate $\chi^2$ per number of degrees of freedom for each of the predictions. This assessment shows clear differences between the predictions obtained with different TMD models, both in the description of the full data set and in the case of individual structure functions. The best description of the structure functions data is obtained with one of the modified WW models. Our analysis can serve to identify the features of the TMD model that should be considered in future gluon TMD fits.}

\tableofcontents

\section{Introduction}

The Drell--Yan (DY) process \cite{Drell:1970wh} is one of the most informative probes of the internal structure of hadrons. In this process, a lepton-antilepton (dilepton) pair is produced through the decay of a neutral electroweak (EW) gauge boson, either a virtual photon $\gamma^*$ or a $Z^0$ boson. The DY $W^\pm$ production can also occur, followed by the decay of the boson into a charged lepton-antineutrino (or charged antilepton-neutrino) pair. The dilepton pair angular distribution is described by functions of the dilepton angles $(\phi, \vartheta)$ and angular coefficients $A_i$ directly related to the DY structure functions (will refer to them as the structure functions interchangeably). They have been extensively studied within the parton model \cite{Collins:1977iv, Kajantie:1978yp, Cleymans:1978ip, Cleymans:1978je}, in which nontrivial relations among them were derived in Ref.\ \cite{Lam:1978pu, Lam:1978zr, Lam:1980uc}.

The structure functions depend on the dilepton invariant mass $M$, its total transverse momentum $q_T$, and its rapidity $y$. For the photon exchange, the only contribution comes from the parity-conserving structure functions, which, including the differential cross section integrated over the dilepton angles, can be parameterized with four independent functions. For the $Z^0$ and $W^\pm$ bosons, there exist additionally five more functions that break parity. The Drell--Yan total cross section and the structure functions were measured accurately at LHC \cite{ATLAS:2014ape, ATLAS:2016rnf}. The structure functions were measured at the $Z^0$ resonance, i.e.\ at $M = M_Z$, where the contribution of $\gamma^*$ is negligible, as functions of the dilepton total transverse momentum $q_T$. Most of the data is well described by the NLO calculations within the collinear perturbative QCD (pQCD) \cite{ATLAS:2016rnf}. Here, order denotes the perturbative order in the calculation of the differential distribution, with the leading contribution occurring at $\mathcal{O}(\alpha_s)$ and corresponding to parton plus EW boson production. One significant exception from the theory and data agreement is the Lam--Tung combination $A_{LT}$, which reaches values around $A_{LT} \approx 0.15$ at $q_T \approx \qty{80}{\GeV}$ while the NLO predictions give values around $A_{LT} \approx 0.09$. The Lam--Tung relation $A_{LT} = 0$ \cite{Lam:1978pu, Lam:1978zr, Lam:1980uc} is satisfied up to the LO ($V^* + \mathrm{jet}$ production) in the collinear pQCD, and its breaking at higher orders is related to real emissions that generate additional transverse momentum causing the rotation of the parton plane with respect to the hadron plane \cite{Faccioli:2011pn, Peng:2015spa}.

The collinear factorization framework for high-energy scattering processes involving hadrons, such as Deep Inelastic Scattering (DIS), Semi-Inclusive DIS (SIDIS), and the Drell--Yan process, is well established and rigorously formulated in pQCD~\cite{Collins:1981tt, Collins:1983ju, Collins:1985ue, Collins:1988ig}. At the leading twist in the collinear factorization framework, the hadronic cross section is given by a convolution of parton distribution functions (PDFs), which encode non-perturbative physics below the factorization scale $\mu_F$, and hard-scattering matrix elements computed in pQCD with on-shell initial partons. Within this formalism, a wide range of observables have been derived and studied, including the dilepton angular distributions in DY production. The DY structure functions were computed within the collinear framework at leading order (LO)~\cite{Kajantie:1978yp, Cleymans:1978ip, Cleymans:1978je, Chaichian:1981va}, next-to-leading order (NLO)~\cite{Mirkes:1992hu, Mirkes:1994dp, Mirkes:1994eb}, and next-to-next-to-leading order (NNLO) accuracy~\cite{Gauld:2017tww} in the perturbative $\alpha_s$ expansion, with the inclusion of NLO EW corrections~\cite{Frederix:2020nyw}. Explicit phenomenological studies exist for $V^*+\mathrm{jet}$ processes~\cite{Pellen:2022fom}, in particular within Monte Carlo implementations~\cite{Alioli:2008gx}. In addition, the impact of resummation effects on the angular decomposition was investigated within the collinear framework~\cite{Boer:2006eq}.

The discrepancy of the NNLO prediction and the $A_{LT}$ data, connected with the fact that the Lam--Tung relation breaks down due to partons transverse momenta, induced the idea of incorporating partons transverse momenta from the beginning. To account for these transverse momenta, we use the so-called high-energy factorization (or $k_T$ factorization), derived and described in \cite{Catani:1990xk, Catani:1990eg}, where, in particular, the formulation of obtaining the cross section from hard matrix elements within this scheme was explained. The DY process within this approach was studied in e.g.\ \cite{Brodsky:1996nj, Kopeliovich:2000fb, Gelis:2002fw, Gelis:2006hy, Schafer:2016qmk, Taels:2023czt}. It was shown in \cite{Motyka:2016lta} that within the $k_T$ factorization approach, significant improvement in the description of $A_{LT}$ can be reached, especially with the Transverse Momentum Distributions (TMDs, alternatively called Transverse Momentum Dependent Parton Distribution Functions --- TMD PDFs) with slow $k_T$ fall-off. This result is the main inspiration for this work. We consider the TMDs used in \cite{Motyka:2016lta} together with other models to check if the $A_{LT}$ description can be improved, and if the remaining structure functions description stays in agreement with data. It should be mentioned that DY structure functions have also been studied within high-energy factorization in an alternative formulation where both quarks and gluons are reggeized~\cite{Nefedov:2012cq, Nefedov:2020ugj, Omelyanchuk:2024wdw, Saleev:2025cgy}. The DY structure functions, with particular emphasis on the Lam--Tung combination, have also been studied within the Color Glass Condensate (CGC) formalism \cite{Gelis:2002fw, Gelis:2006hy}, in particular, within the color-dipole formalism \cite{Bandeira:2025log},  considering also the higher twist effects \cite{Motyka:2014lya, Brzeminski:2016lwh}.

An alternative framework that incorporates partonic transverse momentum effects is the transverse momentum dependent factorization in the Collins--Soper--Sterman (CSS) formalism~\cite{Collins:1984kg}. This approach applies in the region of small transverse momenta, $k_T^2 \ll M^2$, and resums logarithms of the form $\log(M^2/k_T^2)$. In contrast, the $k_T$-factorization is designed for kinematics with $k_T^2 \sim M^2 \ll S$ and resums the large logarithms of $1/x$, employing off-shell partons in the hard scattering matrix elements. The TMD factorization admits a rigorous operator formulation of TMDs and can be systematically matched to the collinear factorization. The DY structure functions in the small dilepton transverse momentum ($q_T$) region have been studied within the CSS framework and compared with experimental data in~\cite{Piloneta:2024aac, Vladimirov:2025qys}.

The parton TMDs parameterize the hadronic structure, providing the basis for QCD evolution in the transverse momentum scale and rapidity. In the small-$x$ regime, the best known evolution equation that resums logarithms of $1/x$ and incorporates transverse momentum is the Balitsky--Fadin--Kuraev--Lipatov (BFKL) equation~\cite{Fadin:1975cb, Kuraev:1976ge, Kuraev:1977fs, Balitsky:1978ic} which is linear and violates unitarity at very small $x$. The Catani--Ciafaloni--Fiorani--Marchesini (CCFM)~\cite{Ciafaloni:1987ur, Catani:1989sg, Catani:1989yc, Marchesini:1994wr} equation interpolates the BFKL evolution at small $x$ and the DGLAP evolution in factorization scale, by taking into account effects of angular ordering and color coherence. At high parton densities, the gluon saturation effects become important and are taken into account in the Balitsky--Kovchegov (BK) equation~\cite{Balitsky:1995ub, Kovchegov:1999yj} and, more generally, by the Jalilian-Marian--Iancu--McLerran--Weigert--Leonidov--Kovner (JIMWLK) evolution equations~\cite{Jalilian-Marian:1997jhx, Jalilian-Marian:1997qno, Weigert:2000gi, Iancu:2000hn, Iancu:2001ad}, which are nonlinear and consistent with unitarity, in contrast with the BFKL equation.

The gluon TMD denoted by $\mathcal{F}\lr{x, \mathbf{k}_T^2, \mu_F^2}$ in general depends on the longitudinal momentum fraction $x$ of the parent hadron, transverse momentum $\mathbf{k}_T$ with respect to hadron momentum, and the factorization scale $\mu_F$. It provides information about the internal structure of hadrons, in particular of the proton. Since it depends on a larger number of parton kinematic variables, it should provide a more accurate description of the hadron structure in the region of applicability of the $k_T$-factorization at a given order in perturbation expansion, than the description given by the collinear PDFs. Therefore, the determination of the accurate TMD parametrization is important for obtaining precise hadron scattering predictions. There are many different TMD models in use today, each with different properties and origins. To distinguish and evaluate these models, it is important to find observables sensitive to transverse momentum effects.

In conclusion, we consider all the Drell--Yan structure functions with a separate consideration of the Lam--Tung combination $A_{LT}$ for different TMD models. In our formalism we take into account two channels --- the scattering of two off-shell gluons $g^* g^* \to q\overline{q}V^*$ and the scattering of the collinear valence quark and off-shell gluon $q_\mathrm{val} g^* \to q V^*$ both at the tree level, where $V^*$ denotes a virtual electroweak gauge boson $\gamma^*$ or $Z^0$. We apply four different types of gluon TMD models, which come from QCD evolution equations or are inspired by QCD. These models are the quasi-collinear Gaussian model which comes from the Golec-Biernat--W\"{u}shoff saturation model \cite{Golec-Biernat:1998zce}, the Jung--Hautmann model which is the result of the CCFM evolution equation \cite{Ciafaloni:1987ur, Catani:1989sg, Catani:1989yc, Marchesini:1994wr}, the Kimber--Martin--Ryskin model \cite{Kimber:1999xc, Kimber:2001sc, Martin:2001ms} which comes from the collinear DGLAP evolution equation \cite{Gribov:1972rt, Lipatov:1974qm, Altarelli:1977zs, Dokshitzer:1977sg} and the Weizs\"{a}cker--Williams model \cite{Motyka:2016lta} which is a phenomenological model based on a concept of the gluon TMD to be driven by the Weizs\"{a}cker--Williams gluon emission from valence quarks, and should not be confused with Weizs\"{a}cker--Williams TMD used in the $k_T$ factorization framework see e.g.\ \cite{Dominguez:2011wm}. We also proposed a modified Weizs\"{a}cker--Williams model which preserves the $k_T$-dependence of the original model, but adopts the dependence on $x$ and the factorization scale from the collinear gluon PDF. For models with the collinear $x$-dependence, we also consider a simple phenomenological modification relying on $x$ rescaling, which should account for the additional generation of invariant mass in the process involving the transverse momenta of partons compared to the collinear process. Some of the models underestimated the total Drell--Yan cross section, which led us to adjust these models by normalizing them to the total cross section data \cite{ATLAS:2014ape}. We observe that the Drell--Yan structure functions indeed differentiate between the gluon TMD models. The Weizs\"{a}cker--Williams model works quite well in comparison to the other models. Its modifications do not significantly change the description of the Lam--Tung combination, but their predictions for other structure functions are different --- some are improved, while others are slightly worsened. We summarize the results in terms of $\chi^2$ per number of degrees of freedom, and their ratio to the minimum $\chi^2$ value.
\section{Kinematics}

We consider two colliding hadrons with momenta $P'_1$ and $P'_2$ that produce a lepton ($l^-$) antilepton ($l^+$) (dilepton) pair. In the standard light-cone coordinates, the hadronic momenta can be written as
\begin{equation}
    P'_1 = \lr{\sqrt{S}, \frac{M_H^2}{\sqrt{S}}; \mathbf{0}} \approx \lr{\sqrt{S}, 0; \mathbf{0}} =: P_1, \qquad P'_2 = \lr{\frac{M_H^2}{\sqrt{S}}, \sqrt{S}; \mathbf{0}} \approx \lr{0, \sqrt{S}; \mathbf{0}} =: P_2,
\end{equation}
where $S = \lr{P_1 + P_2}^2$ is the center of mass collision energy, and we will consistently neglect the hadron mass $M_H$, because in the high energy limit $M_H / \sqrt{S} \ll 1$. The other momenta will be represented using Sudakov decomposition
\begin{equation}
\label{Sudakov decomposition}
    p = x P_1 + \frac{m^2 + \mathbf{p}_T^2}{x S} P_2 + p_T,
\end{equation}
where $p_T = \lr{0, 0; \mathbf{p}_T}$ is such that $p_T \cdot P_1 = p_T \cdot P_2 = 0$ and $m^2 = p^2$.

We denote the virtual boson momentum as $q$, its mass as $M$ and define its transverse mass by $M_T^2 = M^2 + \mathbf{q}_T^2$, so that its Sudakov decomposition reads
\begin{equation}
\label{q decomposition}
    q = x_F P_1 + \frac{M_T^2}{x_F S} P_2 + q_T,
\end{equation}
where $x_F$ is the so called Feynman $x$. The electroweak gauge boson rapidity $y$ in the laboratory frame is related to $x_F$ by the equation $y = \ln(x_F \sqrt{S} / M_T)$ or conversely $x_F = M_T e^y / \sqrt{S}$.

In general we define a coordinate system by the four-vectors $\lr{T, X, Y, Z}$ which are orthogonal and normalized as $T^2 = 1$, $X^2 = Y^2 = Z^2 = -1$. We choose $q$ to define the time direction as $T^\mu \coloneqq q^\mu/M$ and $X$, $Z$ coordinates to span the hadronic plane
\begin{equation}
\label{X, Z general definition}
    X^\mu \coloneqq \alpha_+ \Tilde{P}_{(+)}^\mu + \alpha_- \Tilde{P}_{(-)}^\mu, \qquad Z^\mu \coloneqq \beta_+ \Tilde{P}_{(+)}^\mu + \beta_- \Tilde{P}_{(-)}^\mu,
\end{equation}
where $\alpha_\pm$, $\beta_\pm$ are coefficients that defines the specific frame and we defined the $\pm$ vectors as $P_{(\pm)} \coloneqq P_1 \pm P_2$ with tilde meaning that they are projected onto a hyperplane orthogonal to $q$ i.e.\ $\Tilde{P}_{(\pm)}^\mu \coloneqq \Tilde{g}^{\mu \nu} P_{(\pm) \nu}$ with the projection defined as
\begin{equation}
\label{projection}
    \Tilde{g}^{\mu \nu} \coloneqq g^{\mu \nu} - \frac{q^\mu q^\nu}{M^2}.
\end{equation}
The $Y$ coordinate is defined to complete the right-oriented orthonormal basis.

All the calculations are performed in the Collins--Soper frame \cite{Collins:1977iv}, where the $Z$-axis is defined to bisect the angle formed by the hadrons' momenta directions in the particle rest frame. This definition leaves ambiguity due to two possible ways of choosing bisectors between two non-oriented directions. Such a definition fixes the $Z$-axis up to a sign. The orientation of the $Z$-axis is consistent with the $Z$-direction of the lepton pair in the laboratory reference frame. Therefore, it reverses the sign for the negative rapidity $y$, and due to the requirement of a right-handed basis, the $Y$-axis also reverses the sign. In this frame the coefficients $\alpha_\pm$, $\beta_\pm$ are equal to
\begin{equation}
    \alpha_\pm = \mp \frac{M \lr{q \cdot P_\pm}}{q_T M_T S}, \qquad \beta_\pm = \mp \frac{\lr{q \cdot P_\mp}}{M_T S}.
\end{equation}

We denote the leptons' momenta as $l_1$, $l_2$ for the lepton $l^-$ and antilepton $l^+$ respectively. The angles of the dilepton in the center of momentum (COM) frame are defined through the scalar products with the chosen coordinates
\begin{equation}
\label{lepton angles definition}
    l_{1/2} \cdot X = \mp \frac{M}{2} \sin{\vartheta} \cos{\phi}, \qquad l_{1/2} \cdot Y = \mp \frac{M}{2} \sin{\vartheta} \sin{\phi}, \qquad l_{1/2} \cdot Z = \mp \frac{M}{2} \cos{\vartheta}.
\end{equation}
Given the coordinates, we can define the polarization vectors of the electroweak boson
\begin{equation}
\label{polarization definition}
    \varepsilon_{(0)}^\mu \coloneqq Z^\mu, \qquad \varepsilon_{(\pm)}^\mu \coloneqq \mp \frac{1}{\sqrt{2}} \lr{X^\mu \pm i Y^\mu}.
\end{equation}
\section{The Drell--Yan Process}

In general, the Drell--Yan process is a process that describes lepton pair production in a collision of hadrons $H_1$ and $H_2$. At the leading order in the electromagnetic coupling, the colliding hadrons produce an electroweak gauge boson $V^*$, which then decays into a dilepton pair $l^+ l^-$. It can be written schematically as $H_1\lr{P_1'}+H_2\lr{P_2'}~\to~V^*(q)~+~X(p_X)~\to~l^+(l_2)+l^-(l_1) + X(p_X)$, where $p_X$ denotes a set of $n$ outgoing partons momenta. The factorization formula, assuming the high-energy factorization for the DY cross section $\sigma$, is given by \cite{Collins:1991ty}
\begin{equation}
\label{factorization theorem}
    \dd \sigma = \sum_{i, j} \int \frac{\dd x_1}{x_1} \int \frac{\dd^2 k_{1T}}{\pi} \mathcal{F}_{i/H_1}\lr{x_1, \mathbf{k}_{1T}^2, \mu_F^2} \int \frac{\dd x_2}{x_2} \int \frac{\dd^2 k_{2T}}{\pi} \mathcal{F}_{j/H_2}\lr{x_2, \mathbf{k}_{2T}^2, \mu_F^2} \dd \hat{\sigma}_{ij},
\end{equation}
where $x_{1, 2}$ are the fractions of the longitudinal momentum of the parent hadron carried by a parton, $\mathcal{F}_{i/H}\lr{x, \mathbf{k}_T^2, \mu_F^2}$ is a transverse momentum dependent parton distribution function (TMD) which gives the probability of finding the parton $i$ with longitudinal momentum fraction $x$ and transverse momentum $\mathbf{k}_T$ in the hadron $H$. Parton distributions depend also on the factorization scale $\mu_F$, which separates non-perturbative effects captured by TMDs from the perturbative partonic cross section $\dd \hat{\sigma}_{ij}$. In the case of high-energy factorization, the partonic cross section $\dd \hat{\sigma}_{ij}$ is off-shell while remaining gauge invariant. TMDs are related to the standard collinear parton distribution functions (PDFs) via
\begin{equation}
\label{TMD collinear limit}
    x f_i\lr{x, \mu^2} = \int_0^{\mu^2} \dd k_T^2 \mathcal{F}_i\lr{x, k_T^2, \mu^2}.
\end{equation}

The partonic cross section of a process $i(p_1) j(p_2) \to X(p_X) l^-(l_1) l^+(l_2)$ is given by the formula
\begin{equation}
    \dd \hat{\sigma}_{ij} = \frac{(2 \pi)^4}{F(p_1, p_2)} L_{\mu \nu} \mathcal{M}_{ij}^{\mu \nu} \abs{D_V\lr{q^2}}^2 \dd PS_{n+2}(p_1, p_2; p_X, l_1, l_2),
\end{equation}
where $PS_{n}$ is the phase space of $n$ outgoing particles, $F(p_1, p_2)=\sqrt{(p_1 \cdot p_2)^2 - m_1^2 m_2^2}$, with $m_i^2 = p_i^2$, is the incident flux of the initial partons, $D_V\lr{q^2}$ is the propagator of boson $V$ with mass $M_V$ and decay width $\Gamma_V$ given by $D_V\lr{q^2} = i/\lr{q^2 - M_V^2 + i M_V \Gamma_V}$,
and $L_{\mu \nu}$ is the leptonic tensor
\begin{equation}
\label{leptonic tensor}
    L^{\mu \nu} \coloneqq \sum_{s_1, s_2} \overline{u}_{s_1}(l_1) \Gamma_V^\mu v_{s_2}(l_2) \overline{v}_{s_2}(l_2) \Gamma_V^\nu u_{s_1}(l_1) = 2 g_l^2 \left[ q^\mu q^\nu - l^\mu l^\nu - q^2 g^{\mu \nu} \right] + 2i w_l^2 \varepsilon^{\mu \nu \rho \sigma} q_\rho l_\sigma,
\end{equation}
where the interaction of the boson $V$ with dilepton $l^+ l^-$ is described by the vertex
\begin{equation}
    \Gamma_V^\mu = \lr{v_f^V + a_f^V \gamma_5} \gamma^\mu,
\end{equation}
where $v_f^V$, $a_f^V$ are the vector and axial couplings for the fermion with flavor $f$. We denote leptonic flavors by the index $l$ and quark flavors by the index $q$. We also define $g_l^2 \coloneqq v_l^2 + a_l^2$ and $w_l^2 \coloneqq 2 v_l a_l$ with suppressed boson designation $V$, which is obvious from the context. The formulas for the electroweak couplings are given in Appendix \ref{EW couplings}. The boson momentum is $q = l_1 + l_2$, and we also use the momentum $l^\mu$ given by $l = l_1 - l_2$. We define the partonic amplitude squared with factorized boson polarizations as
\begin{equation}
    \mathcal{M}_{ij}^{\mu \nu} = \overline{\sum} \mathcal{A}^\mu \overline{\mathcal{A}}^\nu,
\end{equation}
where the symbol $\overline{\sum}$ means averaging over the set of indices for the incoming particles and summing over the indices of the outgoing particles and $\mathcal{A}^\mu$ is the scattering amplitude for the process $i(p_1) j(p_2) \to X(p_X) V^*(q)$ with boson polarization removed. The contraction of the leptonic tensor and the above amplitude squared can be written as a sum over the boson basis polarizations
\begin{equation}
    L_{\mu \nu} \mathcal{M}_{ij}^{\mu \nu} = \sum_{r, r' = 0, \pm} L^{(r r')} \mathcal{M}_{ij}^{(r r')},
\end{equation}
with
\begin{equation}
    L^{(r r')} \coloneqq \overline{\varepsilon}^{(r)}_\mu(q) L^{\mu \nu} \varepsilon^{(r')}_\nu(q), \qquad \mathcal{M}_{ij}^{(r r')} \coloneqq \varepsilon^{(r)}_\mu(q) \mathcal{M}_{ij}^{\mu \nu} \overline{\varepsilon}^{(r')}_\nu(q).
\end{equation}

The DY differential cross section can be written in the following form
\begin{equation}
\label{full diff cross sec as polarization sum}
    \frac{\dd \sigma}{\dd M^2 \dd y \dd^2 q_T \dd \Omega} = \frac{4}{(4\pi)^6} \abs{D_V\lr{M^2}}^2 \sum_{r, r' = 0, \pm} L^{(r r')} \frac{\dd \sigma^{(r r')}}{\dd M^2 \dd y \dd^2 q_T},
\end{equation}
where the angular distribution of lepton angles $\Omega = (\vartheta, \phi)$ is given by the polarized leptonic tensor, and we define the polarized differential cross sections as
\begin{equation}
\label{cross section general formula}
\begin{split}
    \frac{\dd \sigma^{(r r')}}{\dd M^2} =& \sum_{i, j} \int \dd x_1 \int \frac{\dd^2 k_{1T}}{\pi x_1} \mathcal{F}_i\lr{x_1, \mathbf{k}_{1T}^2, \mu_F^2} \int \dd x_2 \int \frac{\dd^2 k_{2T}}{\pi x_2} \mathcal{F}_j\lr{x_2, \mathbf{k}_{2T}^2, \mu_F^2} \\
    &\times \frac{(2\pi)^4}{F(p_1, p_2)} \mathcal{M}_{ij}^{(r r')} \dd PS_{n+1}(p_1, p_2; p_X, q).
\end{split}
\end{equation}

\subsection{The Helicity Structure Functions}

The DY differential cross section \eqref{full diff cross sec as polarization sum} can be reduced to the form
\begin{equation}
    \frac{\dd \sigma}{\dd M^2 \dd y \dd^2 q_T \dd \Omega} = \frac{4 g_l^2 M^2}{(4\pi)^6} \abs{D_V\lr{M^2}}^2 \sum_{\tau \in \mathfrak{P}} g_\tau(\vartheta, \phi) \frac{\dd \sigma^\tau}{\dd M^2 \dd y \dd^2 q_T},
\end{equation}
where $\mathfrak{P} \coloneqq \{ U+L, L, TT, LT, A, P, 7, 8, 9 \}$ and with the angular coefficients derived from the leptonic tensor
\begin{align}
    g_{U+L}(\vartheta, \phi) =& 1 + \cos^2{\vartheta}, & g_{L}(\vartheta, \phi) =& 1 - 3 \cos^2{\vartheta}, & g_{TT}(\vartheta, \phi) =& 2 \sin^2{\vartheta} \cos{(2\phi)}, \nonumber \\
    g_{LT}(\vartheta, \phi) =& 2 \sqrt{2} \sin{(2 \vartheta)} \cos{\phi}, & g_{A}(\vartheta, \phi) =& 4\sqrt{2} \sin{\vartheta} \cos{\phi}, & g_{P}(\vartheta, \phi) =& 2 \cos{\vartheta}, \\
    g_7(\vartheta, \phi) =& 2 \sin^2{\vartheta} \sin{(2\phi)}, & g_8(\vartheta, \phi) =& 2\sqrt{2} \sin{(2\vartheta)} \sin{\phi}, & g_9(\vartheta, \phi) =& 4\sqrt{2} \sin{\vartheta} \sin{\phi}, \nonumber
\end{align}
and the helicity differential cross sections $\dd \sigma^\tau$ are
\begin{gather}
    \dd \sigma^{U+L} = \dd \sigma^{(00)} + \dd \sigma^{(++)} + \dd \sigma^{(--)}, \qquad \dd \sigma^{L} \coloneqq \dd \sigma^{(00)}, \qquad \dd \sigma^{TT} = \frac{1}{2} \lr{\dd \sigma^{(+-)} + \dd \sigma^{(-+)}}, \\
    \dd \sigma^{LT} = \frac{1}{4} \lr{\dd \sigma^{(+0)} + \dd \sigma^{(0+)} - \dd \sigma^{(-0)} - \dd \sigma^{(0-)}}, \qquad \dd \sigma^{P} = c_l \lr{\dd \sigma^{(++)} - \dd \sigma^{(--)}}, \\
    \dd \sigma^A = \frac{c_l}{4} \lr{\dd \sigma^{(+0)} + \dd \sigma^{(0+)} + \dd \sigma^{(-0)} + \dd \sigma^{(0-)}}, \qquad \dd \sigma^7 = \frac{1}{2 i} \lr{\dd \sigma^{(+-)} - \dd \sigma^{(-+)}}, \\
    \dd \sigma^8 = \frac{1}{4i} \lr{\dd \sigma^{(+0)} - \dd \sigma^{(0+)} + \dd \sigma^{(-0)} - \dd \sigma^{(0-)}}, \qquad \dd \sigma^9 = \frac{c_l}{4 i} \lr{\dd \sigma^{(+0)} - \dd \sigma^{(0+)} - \dd \sigma^{(-0)} + \dd \sigma^{(0-)}},
\end{gather}
where $c_l \coloneqq w_l^2 / g_l^2 = 2 v_l a_l / \lr{v_l^2 + a_l^2}$.

The differential cross section can then be written as
\begin{equation}
\begin{split}
    \frac{\dd \sigma}{\dd M^2 \dd y \dd^2 q_T \dd \Omega} = \frac{3}{16 \pi} \frac{\dd \sigma}{\dd M^2 \dd y \dd^2 q_T} \bigg[& 1+\cos^2{\vartheta} + \frac{1}{2} A_0 \lr{1-3\cos^2{\vartheta}} + A_1 \sin{(2 \vartheta)} \cos{\phi} + \\
    &+ \frac{1}{2} A_2 \sin^2{\vartheta} \cos{(2 \phi)} + A_3 \sin{\vartheta} \cos{\phi} + A_4 \cos{\vartheta} + \\
    &+ A_5 \sin^2{\vartheta} \sin{(2 \phi)} + A_6 \sin{(2\vartheta)} \sin{\phi} + A_7 \sin{\vartheta} \sin{\phi} \bigg],
\end{split}
\end{equation}
where the differential cross section integrated over the lepton angles is
\begin{equation}
    \frac{\dd \sigma}{\dd M^2 \dd y \dd^2 q_T} = \frac{16 \pi}{3} \frac{4 g_l^2 M^2}{(4\pi)^6} \abs{D_V\lr{M^2}}^2 \frac{\dd \sigma^{U+L}}{\dd M^2 \dd y \dd^2 q_T},
\end{equation}
and the $A_i$ structure functions are defined as follows
\begin{equation}
\label{structure functions definition}
\begin{split}
    A_0 = \frac{2 W_L}{W_{U+L}}, \qquad A_1 = \frac{2 \sqrt{2} W_{LT}}{W_{U+L}}, \qquad A_2 =& \frac{4 W_{TT}}{W_{U+L}}, \qquad A_3 = \frac{4 \sqrt{2} W_A}{W_{U+L}}, \qquad A_4 = \frac{2 W_P}{W_{U+L}}, \\
    A_5 = \frac{2 W_7}{W_{U+L}}, \qquad A_6 =& \frac{2 \sqrt{2} W_8}{W_{U+L}}, \qquad A_7 = \frac{4 \sqrt{2} W_9}{W_{U+L}},
\end{split}
\end{equation}
with the designation $W_\tau = \cfrac{\dd \sigma^\tau}{\dd M^2 \dd y \dd^2 q_T}$.

The structure functions obey the so called Lam--Tung relation \cite{Lam:1978pu, Lam:1980uc}
\begin{equation}
\label{Lam-Tung Relation}
    A_{LT} \coloneqq A_0 - A_2 = 0,
\end{equation}
which holds up to the NLO accuracy in the collinear factorization \cite{Peng:2015spa}.

\section{The Drell--Yan Cross Sections}

We adopt the approximation concerning the hierarchy of the scattering processes following Ref.~\cite{Motyka:2016lta} ---- we assume that the sea quarks entering the hard scattering are generated either directly in the matrix element or in the last step of the parton evolution via gluon splitting. In this approach we consider two partonic channels, namely $q_\mathrm{val} g^*$ channel with polarized cross sections $\dd \sigma^{(r_1 r_2)}_{(qg^* + g^*q)} \coloneqq \dd \sigma^{(r_1 r_2)}_{(qg^*)} + \dd \sigma^{(r_1 r_2)}_{(g^*q)}$ and $g^* g^*$ channel with polarized cross sections $\dd \sigma^{(r_1 r_2)}_{(g^*g^*)}$. The total cross section is a sum of the cross sections for these two channels
\begin{equation}
\label{total cross section}
    \dd \sigma^{(r_1 r_2)} = \dd \sigma^{(r_1 r_2)}_{(qg^*+g^*q)} + \dd \sigma^{(r_1 r_2)}_{(g^*g^*)}.
\end{equation}
Moreover, we are using the so-called nonsense polarization \cite{Collins:1991ty} for the initial off-shell gluons
\begin{equation}
\label{nonsense polarization}
    \varepsilon^\mu(k_i) = \frac{x_i}{\sqrt{-k_{iT}^2}} P_i^\mu,
\end{equation}
where the gluons momenta are decomposed as $k_i = x_i P_i + k_{iT}$ for $i=1, 2$.

\subsection{Cross Section for \texorpdfstring{$q_\mathrm{val} g^*$}{q\mathrm{val} g*} Channel}

For the $q_\mathrm{val} g^*$ channel, depicted in Figure \ref{qg general diagram}, we are using the so-called hybrid factorization, where the initial gluon $g^*$ is off-shell and carries the transverse momentum $k_T$ while the valence quark $q_\mathrm{val}$ is assumed to be collinear. This contribution was derived earlier within the $k_T$-factorization approach in Ref.\ \cite{Brodsky:1996nj, Kopeliovich:2000fb, Gelis:2002fw}. According to \eqref{cross section general formula}, taking the collinear limit \eqref{TMD collinear limit} for the valence quark, we obtain the polarized cross section for this channel expressed as
\begin{equation}
\label{differential cross section qG}
    \frac{\dd \sigma^{(r_1 r_2)}_{(qg^*)}}{\dd M^2 \dd y \dd^2 q_T} = \sum_q \int \dd x_q f_{q, \text{val}}(x_q, \mu_F) \int \frac{\dd^2 k_T}{\pi \mathbf{k}_T^2} \mathcal{F}\lr{x_g, \mathbf{k}_T^2, \mu_F^2} \frac{4 \pi \alpha_s(\mu_F)}{N} \frac{z^2}{(8 \pi)^2 x_F^2 (1 - z) S^2} \Phi_{r_1 r_2}^{(q)},
\end{equation}
where $f_{q, \text{val}}$ is a PDF for a valence quark $q$, the gluon momentum fraction $x_g$ is fixed from the energy-momentum conservation as $x_g = \left( (1 - z) M_T^2 + z \lr{\mathbf{k}_T - \mathbf{q}_T}^2 \right)/\left( x_F (1 - z) S \right)$ with $z = x_F/x_q$ and the impact factor is defined as
\begin{equation}
\label{Impact Factor qG}
    \Phi_{r_1 r_2}^{(q)} \coloneqq \sum_{\sigma_1, \sigma_2} A_{\sigma_1 \sigma_2}^{(r_1)} \overline{A}_{\sigma_1 \sigma_2}^{(r_2)},
\end{equation}
where $A_{\sigma_1 \sigma_2}^{(r_1)} = \varepsilon_{(r_1), \mu} A_{\sigma_1 \sigma_2}^\mu$ also carries the information about the quark flavor $q$. The scattering amplitudes for this channel were derived in the Appendix \ref{amplitudes qg derivation}. The $\dd \sigma^{(r_1 r_2)}_{(g^* q)}$ contribution can be obtained by swapping the proton beams.

\subsection{Cross Section for \texorpdfstring{$g^* g^*$}{g* g*} Channel}

\begin{figure*}[t!]
    \begin{subfigure}[t]{0.49\textwidth}
        \centering
        \includegraphics[height=7cm]{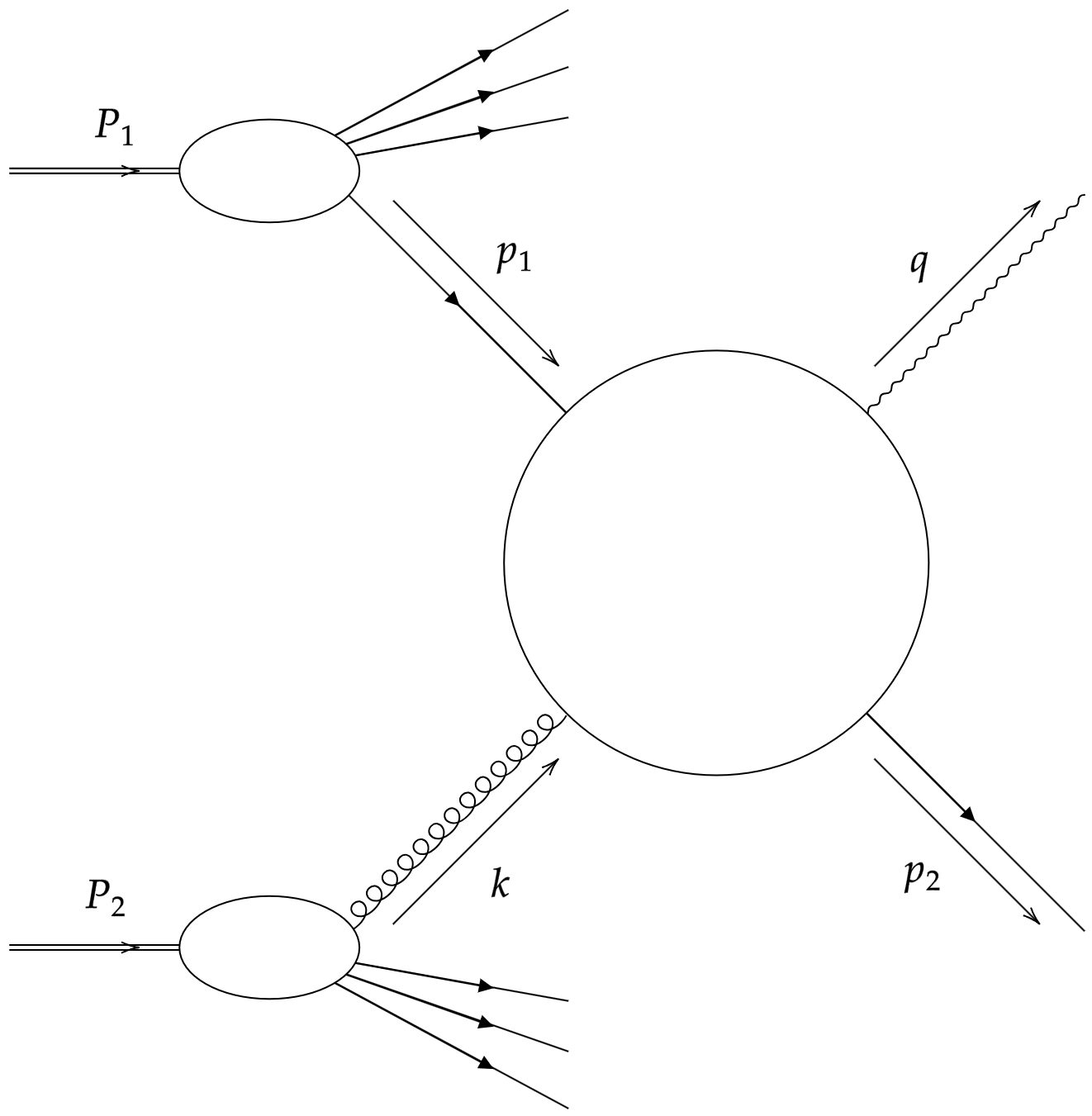}
        \caption{Diagram for the subprocess $q_\mathrm{val} g^* \to q V^*$, where $p_1$ is the momentum of the incoming valence quark, $k$ is the momentum of the off-shell incoming gluon and $p_2$ is the momentum of the outgoing quark.}
        \label{qg general diagram}
    \end{subfigure}
    ~
    \begin{subfigure}[t]{0.49\textwidth}
        \centering
        \includegraphics[height=7cm]{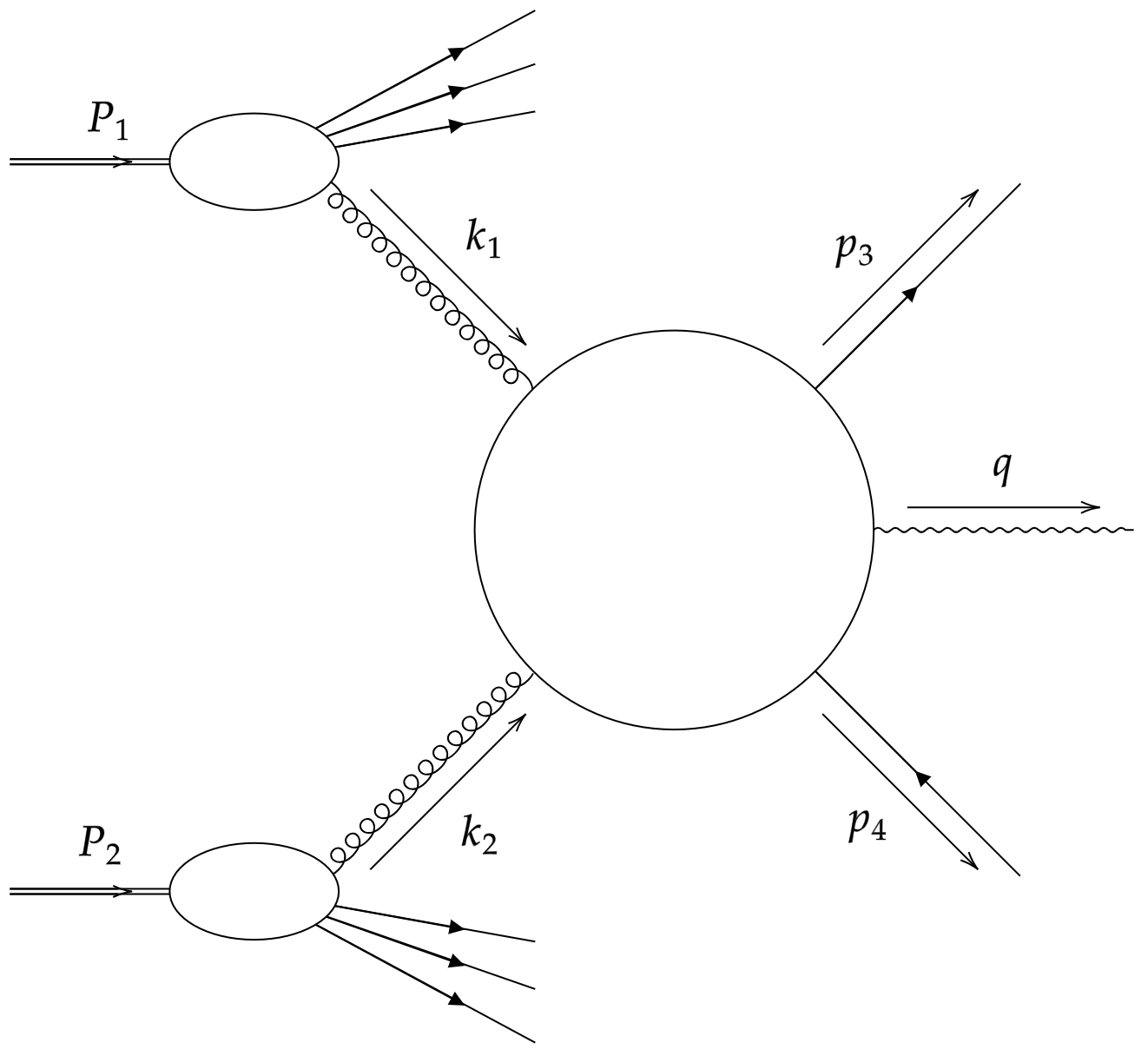}
        \caption{Diagram for the subprocess $g^* g^* \to q\overline{q} V^*$, where $k_1$, $k_2$ are the momenta of two off-shell gluons coming from each hadron and $p_3$, $p_4$ are the momenta of the outgoing quark and antiquark, respectively.}
        \label{gg general diagram}
    \end{subfigure}
    \caption[font=12pt]{General form of the diagrams for the DY process. Big circles represent all possible QCD LO subdiagrams. $P_1$, $P_2$ are the momenta of the incoming hadrons and $q$ is the momentum of the emitted electroweak gauge boson.}
    \label{general diagrams}
\end{figure*}

In the $g^* g^*$ channel, we consider the scattering of two off-shell gluons at the tree level $g^* g^* \to q \overline{q} V^*$. The general form of a diagram for such a process is illustrated in Figure \ref{gg general diagram}.

The contribution for this channel was calculated earlier, e.g.\ \cite{Motyka:2016lta, Deak:2008ky}. From formula \eqref{cross section general formula}, we obtain the cross section for this channel in the form
\begin{equation}
\begin{split}
\label{differential cross section GG}
    \dd \sigma^{(r_1 r_2)}_{(g^* g^*)} =& \int \dd x_1 \int \frac{\dd^2 k_{1T}}{\pi \mathbf{k}_{1T}^2} \mathcal{F}\lr{x_1, \mathbf{k}_{1T}^2, \mu_F^2} \int \dd x_2 \int \frac{\dd^2 k_{2T}}{\pi \mathbf{k}_{2T}^2} \mathcal{F}\lr{x_2, \mathbf{k}_{2T}^2, \mu_F^2} \\
    &\times \frac{(2 \pi)^4}{2 S} \mathcal{M}_{(g^* g^*)}^{(r_1 r_2)} \dd PS_3(k_1, k_2; p_3, p_4, q),
\end{split}
\end{equation}
where $\mathcal{M}_{(g^* g^*)}^{(r_1 r_2)}$ is polarized amplitude squared (with pulled out $x_i^2/\mathbf{k}_{iT}^2$ factors from gluon polarizations), which is given by \eqref{Amplitude squared gg} and $\dd PS_3$ is a three-body phase space
\begin{equation*}
    \dd PS_3(k_1, k_2; p_3, p_4, q) = \frac{\dd z \dd \phi_\kappa \dd y \dd^2 q_T}{8 (2 \pi)^9}.
\end{equation*}
The chosen integration variables $z, \phi_\kappa$ are defined through the quark and antiquark momenta $p_3$ and $p_4$ decomposition
\begin{equation}
\label{massless quarks decomposition}
    p_3 = z x_{q\overline{q}} P_1 + \frac{\mathbf{p}_3^2}{z x_{q\overline{q}} S} P_2 + p_{3T}, \qquad p_4 = (1 - z) x_{q\overline{q}} P_1 + \frac{\mathbf{p}_4^2}{(1 - z) x_{q\overline{q}} S} P_2 + p_{4T},
\end{equation}
where $\mathbf{p}_3$ and $\mathbf{p}_4$ are transverse momenta which can be written in terms of the COM transverse momentum $\bm \Delta$ and momentum $\bm \kappa = \kappa \lr{\cos{\phi_\kappa}, \sin{\phi_\kappa}}$ as
\begin{equation}
    \mathbf{p}_3 = z \bm \Delta + \bm \kappa, \qquad \mathbf{p}_4 = (1 - z) \bm \Delta - \bm \kappa.
\end{equation}
By the energy-momentum conservation Dirac deltas, we get the following constraints
\begin{equation}
\begin{split}
    x_{q\overline{q}} = x_1 - x_F, \qquad \bm \Delta = \mathbf{k}_{1T} + \mathbf{k}_{1T} - \mathbf{q}_T, \qquad \kappa^2 = z (1 - z) \left( x_2 x_{q\overline{q}} S - \frac{x_{q\overline{q}} M_T^2}{x_F} - \bm \Delta^2 \right).
\end{split}
\end{equation}
\section{Models for Gluon TMD}
\label{Models for Gluon TMD}

We consider four models of gluon TMD proposed earlier, and we introduce some modifications to them. The models were applied earlier in calculations of the DY structure functions measured by the ATLAS collaboration \cite{ATLAS:2016rnf}.

Thus, we consider the following TMD models:
\begin{itemize}
    \item the Gaussian (Gauss.) model,
    \item the Jung--Hautmann (JH) model \cite{Hautmann:2013tba},
    \item the Kimber--Martin--Ryskin (KMR) model \cite{Kimber:2001sc},
    \item the ``Weizs\"{a}cker--Williams'' (WW) model \cite{Motyka:2016lta}.
\end{itemize}
We will briefly summarize their main characteristics, underlying ideas, and the proposed modifications, along with their justification.

\subsection*{The Gaussian Model}

The Gaussian gluon distribution is inspired by the Golec-Biernat--W\"{u}sthoff (GBW) saturation model \cite{Golec-Biernat:1999qor} and extracted from the color-dipole cross-section. It is referred to as quasi-collinear due to its very narrow $k_T$ dependence of the form
\begin{equation}
    \mathcal{F}_{\mathrm{G}}\!\left( x, k_T^2 \right) = \frac{N_2}{k_0^2} (1 - x)^7 \exp[-\left(\frac{x}{x_0} \right)^\lambda \frac{k_T^2}{k_0^2}],
\end{equation}
where we take the values of parameters used in Ref.\ \cite{Motyka:2016lta}, $N_2 = 68.4$, $k_0 = \qty{1}{\GeV}$, $x_0 = 3\cdot10^{-4}$ and $\lambda = 0.29$. The parameters $\lambda$ and $x_0$ were fitted to inclusive DIS data on $F_2$ \cite{Golec-Biernat:1998zce} and normalization was adjusted in Ref.\ \cite{Motyka:2016lta}. It provides a simple, phenomenological description of the gluon density that captures saturation effects at small $x$, although it neglects QCD evolution with the factorization scale other than gluon $k_T$.

\subsection*{The Jung--Hautmann Model}

The JH TMD is obtained from the CCFM evolution equation \cite{Ciafaloni:1987ur, Catani:1989sg, Catani:1989yc}, which attempts to unify the small-$x$ BFKL evolution equation \cite{Fadin:1975cb, Kuraev:1976ge, Kuraev:1977fs, Balitsky:1978ic} and the collinear DGLAP evolution \cite{Dokshitzer:1977sg, Altarelli:1977zs}, thus providing a consistent treatment of both $\log(1/x)$ and the logarithm in the hard scale.

Currently, there are a few available grids in the \texttt{TMDlib} library that differ from each other. We chose to compare three sets of gluon TMD grids --- \texttt{JH-2013-set1} \cite{Hautmann:2014kza}, \texttt{PB-NLO-HERAI+II-2018-set1} \cite{BermudezMartinez:2018fsv}, and \texttt{PB-NLO-HERAI+II-2023-set2-qs=1.04} \cite{Bubanja:2023nrd}, which we will call JH2013, PB2018, and PB2023, respectively, and we will collectively call them the JH models. All mentioned grids were obtained using the same Monte Carlo method for solving the evolution equation. However, the evolution kernel used for the JH2013 was dominated by the gluon channel, while the kernel used for the two remaining grids has a fully coupled flavor structure \cite{Hautmann:2017xtx}.

In order to obtain intermediate TMD values from the grid files, we applied linear interpolation. One should be aware of the limited range of the \texttt{JH-2013-set1} grid i.e.\ the maximum value for the transverse momentum is $k_T = \qty{200}{\GeV}$.

\subsection*{The Kimber--Martin--Ryskin Model}

The KMR model was derived by resumming the virtual contribution from the DGLAP equation by including the last splitting as non-integrated. The KMR distribution takes the form
\begin{equation}
\label{differential KMR TMD}
    \mathcal{F}_{\mathrm{KMR}}\lr{x, k_T^2, \mu^2} = \frac{\partial}{\partial k_T^2} \left[ x f_g\lr{x, k_T^2} T_g\lr{k_T, \mu} \right],
\end{equation}
where $f_g$ is the collinear gluon PDF and
\begin{equation}
\label{Sudakov form factor}
    T_g(k_T, \mu) = \exp[-\int_{k_T^2}^{\mu^2} \frac{\dd p^2}{p^2} \frac{\alpha_s(p^2)}{2 \pi} \int_0^{1 - \delta(p)} \dd z z \left( P_{gg}(z) + \sum_q P_{qg}(z) \right)],
\end{equation}
is the Sudakov form factor that gives the probability of the gluon with transverse momentum $\mathbf{k}_T$ remaining unchanged by the evolution up to the factorization scale $\mu$. The functions $P_{ij}$ are the first order Altarelli--Parisi splitting functions of the form
\begin{subequations}
\label{splitting functions}
\begin{gather}
    P_{gg}(z) = 2 C_A \frac{\lr{1 - z (1 - z)}^2}{1 - z}, \label{gg splitting} \\
    P_{qg}(z) = \frac{1}{2} \lr{z^2 + (1 - z)^2}, \label{qg splitting} \\
    P_{gq}(z) = C_F \frac{1 + (1-z)^2}{z}, \label{gq splitting}
\end{gather}
\end{subequations}
and $\delta(p) = p/(p+\mu)$ is the parameter that provides the correct angular ordering of the real gluon momenta. The form factor is set to $T_g(Q, \mu) = 1$ for $Q > \mu$. Because of that, the differential form of the KMR TMD \eqref{differential KMR TMD} has a discontinuity due to non-differentiability of the differentiated function at $Q = \mu$. However, one can eliminate this problem by using the integral form \cite{Golec-Biernat:2018hqo}
\begin{equation}
\label{integral KMR TMD}
    \mathcal{F}_{\mathrm{KMR}}\lr{x, k_T^2, \mu^2} = \frac{T_g(k_T, \mu)}{k_T^2} \frac{\alpha_s\lr{k_T^2}}{2 \pi} \int_x^{1-\delta\lr{k_T}} \dd z \left[ P_{gg}(z) \frac{x}{z} f_g\lr{\frac{x}{z}, k_T^2} + \sum_q P_{gq}(z) \frac{x}{z} f_q\lr{\frac{x}{z}, k_T^2} \right],
\end{equation}
where the sum $\sum_q$ runs over all active quark and antiquark flavors. Equations \eqref{differential KMR TMD} and \eqref{integral KMR TMD} agree for $k_T < \mu$ and start to deviate for $k_T > \mu$. This behavior is due to the use of a regularized gluon PDF and can be avoided by using a cut-off dependent DGLAP solution for the gluon PDF in the TMD definition \eqref{differential KMR TMD} \cite{Golec-Biernat:2018hqo}. However, the available grids for PDFs are cut-off free ($\delta(k_T) = 0$), therefore we use formula \eqref{integral KMR TMD} in our computations.

For the computations, we generated a grid for the KMR model using formula \eqref{integral KMR TMD} and we defined the function as the linear interpolation on this grid. As the input PDF we employed the \texttt{CT10nlo} model \cite{Lai:2010vv} from the \texttt{LHAPDF} library, together with the one-loop running coupling $\alpha_s\lr{\mu^2}$ given by
\begin{equation}
\label{1-loop coupling}
    \alpha_s\lr{\mu^2} = \frac{1}{\beta_f \log(\frac{\mu^2}{\Lambda_f^2})},
\end{equation}
where $\beta_f = \frac{33 - 2 n_f}{12 \pi}$ for $n_f$ being the number of active flavors and $\Lambda_f$ is the QCD scale, which also depends on the number of active quark flavors $n_f$. The grid covers the ranges $x \in [\qty{1e-7}, 1]$, $k_T \in [\qty{1e-1}{\GeV}, \qty{1e4}{\GeV}]$ and $\mu \in [\qty{2}{\GeV}, \qty{1024}{\GeV}]$. Below a certain scale $k_0$ that parametrizes the onset of non-perturbative physics, which we chose to be $k_0 = \qty{1}{\GeV}$, we fix the TMD to be constant in $k_T$. Thus, for $k_T < k_0$, we set the TMD to be
\begin{equation}
    \mathcal{F}_{\mathrm{KMR}}\lr{x, k_T^2, \mu^2} = \frac{x f_g\lr{x, k_0^2}}{k_0^2} T_g\lr{k_0, \mu},
\end{equation}
to preserve the collinear limit.

We also considered modifications of this model. Initially, we observed that it did not reproduce the correct values for the total DY cross section with the virtual photon exchange measured by ATLAS \cite{ATLAS:2014ape}. Therefore, as an adjustment, we rescaled the TMD by the normalization constant fitted to the total DY cross section data.

As one can see from equation \eqref{differential KMR TMD}, the $x$-dependence of the TMD is governed by that of the collinear PDF. However, the collinear and transverse-momentum-dependent approaches are characterized by different kinematics. As a consequence, the nonzero $\mathbf{k}_T$ gluon momentum fraction $x$ has to be larger than the collinear gluon $x$ to reproduce the same invariant mass of the final state. Therefore, we introduce a simple phenomenological correction by rescaling $x$ in the TMD by a constant factor. Together with an unscaled $x$, we considered three cases:
\begin{subequations}
\label{Modified KMR}
\begin{gather}
    \mathcal{F}^{(1)}_{\mathrm{KMR}}\lr{x, k_T^2, \mu^2} = N^{(1)}_{\mathrm{KMR}} \mathcal{F}_{\mathrm{KMR}}\lr{x, k_T^2, \mu^2}, \label{Modified KMR 1}\\
    \mathcal{F}^{(2)}_{\mathrm{KMR}}\lr{x, k_T^2, \mu^2} = N^{(2)}_{\mathrm{KMR}} \mathcal{F}_{\mathrm{KMR}}\lr{x/2, k_T^2, \mu^2}, \label{Modified KMR 2}\\
    \mathcal{F}^{(3)}_{\mathrm{KMR}}\lr{x, k_T^2, \mu^2} = N^{(3)}_{\mathrm{KMR}} \mathcal{F}_{\mathrm{KMR}}\lr{x/4, k_T^2, \mu^2}, \label{Modified KMR 3}
\end{gather}
\end{subequations}
where each case was additionally rescaled by the overall normalization constant $N^{(i)}_{\mathrm{KMR}}$ fitted to the total DY total cross section data \cite{ATLAS:2014ape}. This should be understood as building a data-guided phenomenological model. This approach looses unfortunately the QCD direct connection to collinear gluon PDF. We will denote the $\mathcal{F}^{(i)}_{\mathrm{KMR}}$ TMDs as $\text{KMR}^{(i)}$.

\subsection*{The Weizs\"{a}cker--Williams Model}

The simple WW model proposed in \cite{Motyka:2016lta} is given by the formula
\begin{equation}
\label{WW TMD}
    \mathcal{F}_{\mathrm{WW}}\lr{x, k_T^2} = \frac{N_1}{k_0^2} (1 - x)^7 x^{-\lambda b} \times \begin{dcases}
        1 & k_T^2 < k_0^2 \\
        \left( \frac{k_0^2}{k_T^2} \right)^b & k_T^2 \geq k_0^2
    \end{dcases},
\end{equation}
where $N_1 = 0.6111$, $\lambda = 0.29$, $b = 1$ and $k_0 = \qty{1}{\GeV}$. The normalization constant $N_1$ is obtained by fitting the DY cross section with a global $K$-factor for photon exchange to the data \cite{ATLAS:2014ape}. The transverse momentum cut-off $k_0$ is chosen as the scale that delimits the region where the confinement or parton coherence effects in a hadron become important.

To capture the collinear limit of TMD \eqref{TMD collinear limit}, we propose the modification of the Weizs\"{a}cker--Williams model (WW'), which uses the collinear gluon PDF. The formula for WW' TMD reads (for $b = 1$)
\begin{equation}
\label{Modified WW TMD}
    \mathcal{F}_{\mathrm{WW}}'\lr{x, k_T^2, \mu^2} = \frac{x f_g\left( x, \mu^2 \right)}{k_0^2 \left[ 1 + \log(\frac{\mu^2}{k_0^2}) \right]} \times \begin{dcases}
        1 & k_T^2 < k_0^2 \\
        \frac{k_0^2}{k_T^2} & k_T^2 \geq k_0^2
    \end{dcases}.
\end{equation}
This modification preserves the $1/k_T^2$ behavior of the original WW model and satisfies equation \eqref{TMD collinear limit}. Additionally, it introduces a double scale dependence present in the JH and KMR models.

Although we did not use it, the values of parameter $b$ other than $1$ have been explored before. For $b \neq 1$, the WW' model takes the form
\begin{equation}
\label{Modified WW TMD b}
    \mathcal{F}_{\mathrm{WW}}'\lr{x, k_T^2, \mu^2} = \frac{(1 - b) x f_g\left( x, \mu^2 \right)}{\lr{\frac{k_0^2}{\mu^2}}^b \mu^2 - b k_0^2} \times \begin{dcases}
        1 & k_T^2 < k_0^2 \\
        \lr{\frac{k_0^2}{k_T^2}}^b & k_T^2 \geq k_0^2
    \end{dcases},
\end{equation}
which reduces to \eqref{Modified WW TMD} in the $b \to 1$ limit.

As in the case of the KMR model, the WW' model also does not describe well the data from Ref.~\cite{ATLAS:2014ape}. Therefore, we consider modifications analogous to \eqref{Modified KMR} for the WW' model
\begin{subequations}
\label{Rescaled Modified WW TMD}
\begin{gather}
    \mathcal{F}^{(1)}_{\mathrm{WW}}\lr{x, k_T^2, \mu^2} = N^{(1)}_{\mathrm{WW}} \mathcal{F}'_{\mathrm{WW}}\lr{x, k_T^2, \mu^2}, \label{Modified WW 1}\\
    \mathcal{F}^{(2)}_{\mathrm{WW}}\lr{x, k_T^2, \mu^2} = N^{(2)}_{\mathrm{WW}} \mathcal{F}'_{\mathrm{WW}}\lr{x/2, k_T^2, \mu^2}, \label{Modified WW 2}\\
    \mathcal{F}^{(3)}_{\mathrm{WW}}\lr{x, k_T^2, \mu^2} = N^{(3)}_{\mathrm{WW}} \mathcal{F}'_{\mathrm{WW}}\lr{x/4, k_T^2, \mu^2}, \label{Modified WW 3}
\end{gather}
\end{subequations}
where again $N^{(i)}_{\mathrm{WW}}$ were fitted to the total DY cross section. We will denote the $\mathcal{F}^{(i)}_{\mathrm{WW}}$ TMDs as $\text{WW}^{(i)}$.

As a collinear gluon PDF, we used the \texttt{CT10nlo} set from the \texttt{LHAPDF} library.

\section{Results}
\label{Results}

To date, the best measurement of the Drell--Yan dilepton angular coefficients was performed by the ATLAS collaboration \cite{ATLAS:2016rnf}, which is the main source of the data on which we base our analysis. We derived the structure functions within the $k_T$ factorization formalism for the $q_\mathrm{val} g^*$ and $g^* g^*$ channels. We focused on the first three parity-conserving functions $A_0$, $A_1$, and $A_2$, considering separately the Lam--Tung combination $A_{LT} = A_0 - A_2$, and two parity-violating functions $A_3$ and $A_4$. The remaining three structure functions --- $A_5$, $A_6$, and $A_7$ --- vanish within the framework of the considered model. The data show that these functions are indeed consistent with zero at current accuracy. In definitions \eqref{structure functions definition} we used the differential cross sections \eqref{differential cross section qG} and \eqref{differential cross section GG} fully integrated over all variables except the boson transverse momentum $q_T$ over which we take the average in the bins used in experiment and in the $Z^0$ boson peak i.e.\ at $M = M_Z = \qty{91.1876}{\GeV}$ at the $pp$ collision energy $\sqrt{S} = \qty{8}{\TeV}$. In our analysis, we considered four main gluon TMD models along with modifications of two of them. The gluon TMD models used are listed and briefly described in Section \ref{Models for Gluon TMD}. For a collinear valence quark PDF in \eqref{differential cross section qG} we used the \texttt{CT10nlo} set from the \texttt{LHAPDF} library. The phase space integration is performed using the \texttt{vegas 6.2.1} package \cite{Lepage:2020tgj}. For a running strong coupling we used one-loop coupling \eqref{1-loop coupling} for the number of flavors $n_f = 5$ where the QCD scale has the value $\Lambda_5 = \qty{98.61}{\MeV}$ that gives $\alpha_s\lr{M_Z^2} = 0.12$. The factorization and renormalization scales are set to be the same and equal to the boson transverse mass $\mu_F = \mu_R = M_T$. To obtain the electroweak couplings, we need the Weinberg mixing angle $\theta_W$ for which we took the value $\sin^2{\theta_W} = 0.23122$ \cite{ParticleDataGroup:2024cfk}.

\subsection{Normalization to the Total Drell--Yan Cross Section}
\label{Results DY CS}

The basic versions of the gluon TMD models do not accurately describe the DY total cross section data given in Ref. \cite{ATLAS:2014ape}, mostly because of the inaccurate overall normalization. Therefore, as a simple phenomenological adjustment, we normalized these functions to the data. In the case of the KMR model \eqref{integral KMR TMD} and WW' model \eqref{Modified WW TMD}, the $x$ dependence is based on the collinear PDFs. However, as mentioned in the previous section, different kinematics in the $k_T$ factorization approach may result in a change in the shape of the distributions in the $x$ variable. Thus, we also proposed versions of the discussed TMDs with the rescaled $x$ variable. Ultimately, we consider the KMR-based TMD family given in \eqref{Modified KMR}, and the WW'-based TMD family given in \eqref{Rescaled Modified WW TMD}, which together with the original WW TMD \eqref{WW TMD} will be referred to as WW-based TMDs.

\begin{figure*}[h!]
    \begin{subfigure}[h]{0.49\textwidth}
        \centering
        \includegraphics[width=1\textwidth]{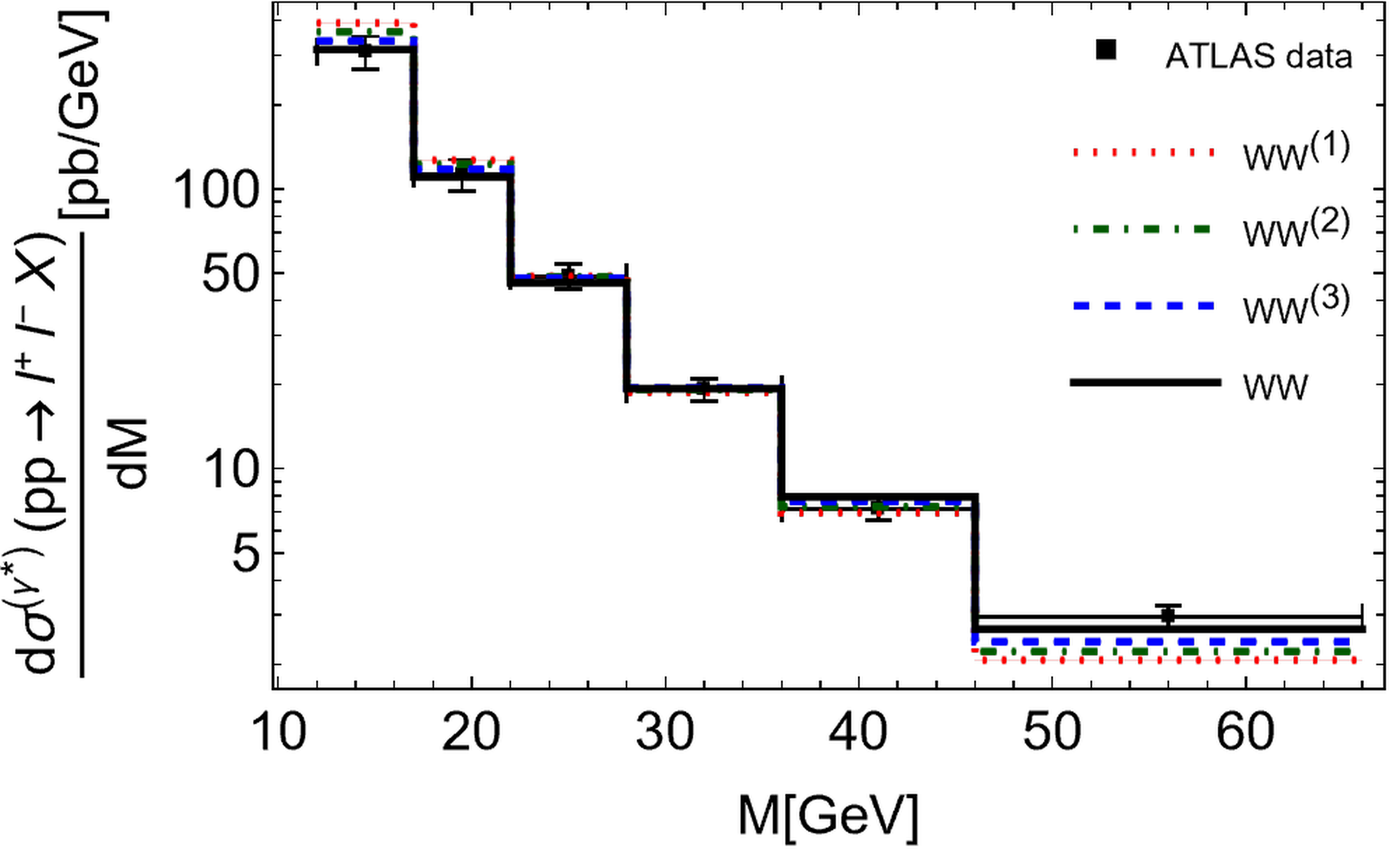}
        \caption{$\dd \sigma^{(\gamma^*)}/\dd M$ for WW-based TMD models.}
        \label{CS WW plot}
    \end{subfigure}
    ~
    \begin{subfigure}[h]{0.49\textwidth}
        \centering
        \includegraphics[width=1\textwidth]{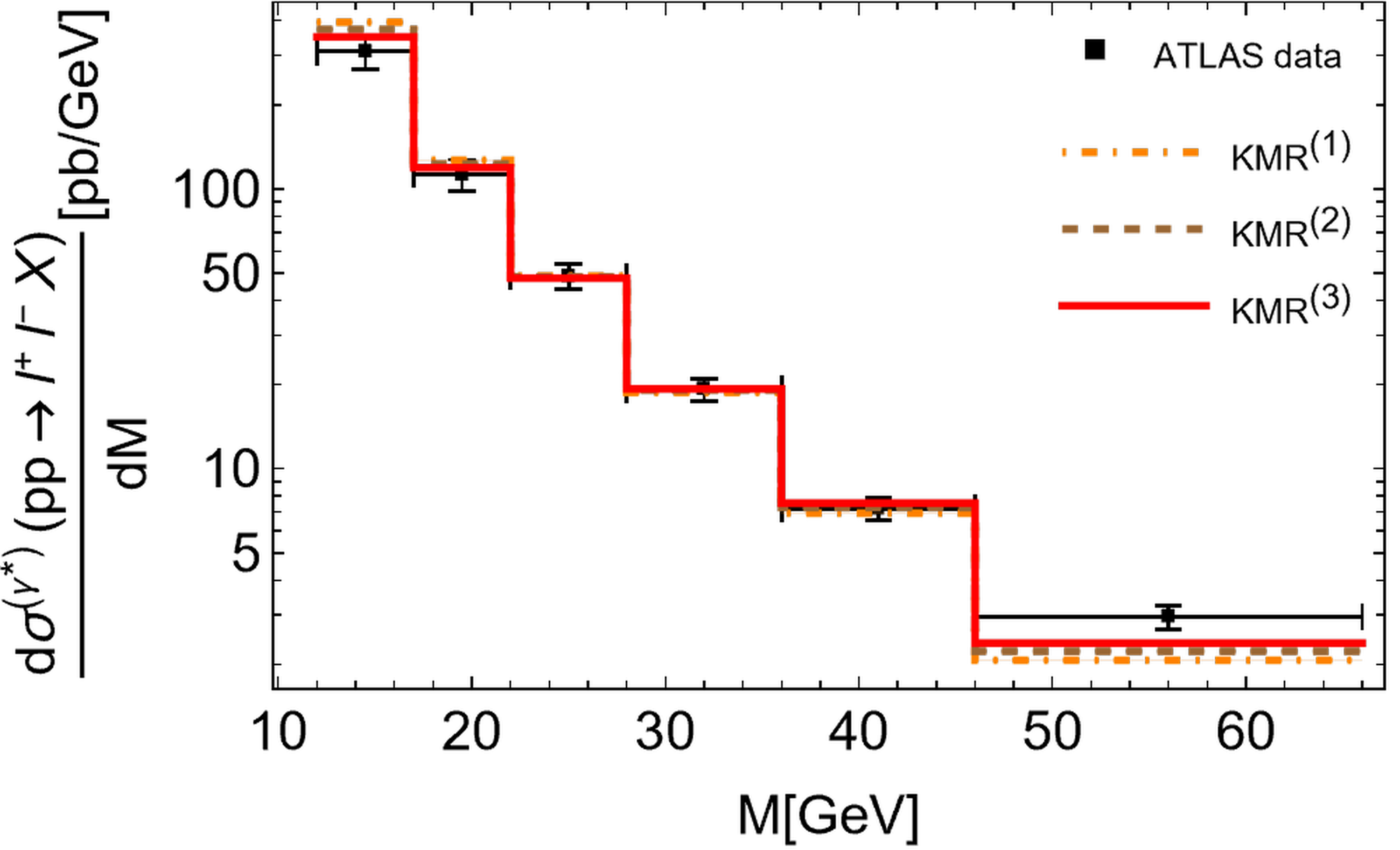}
        \caption{$\dd \sigma^{(\gamma^*)}/\dd M$ for KMR-based TMD models}
        \label{CS KMR plot}
    \end{subfigure}
    ~
    \begin{subfigure}[h]{0.49\textwidth}
        \centering
        \includegraphics[width=1\textwidth]{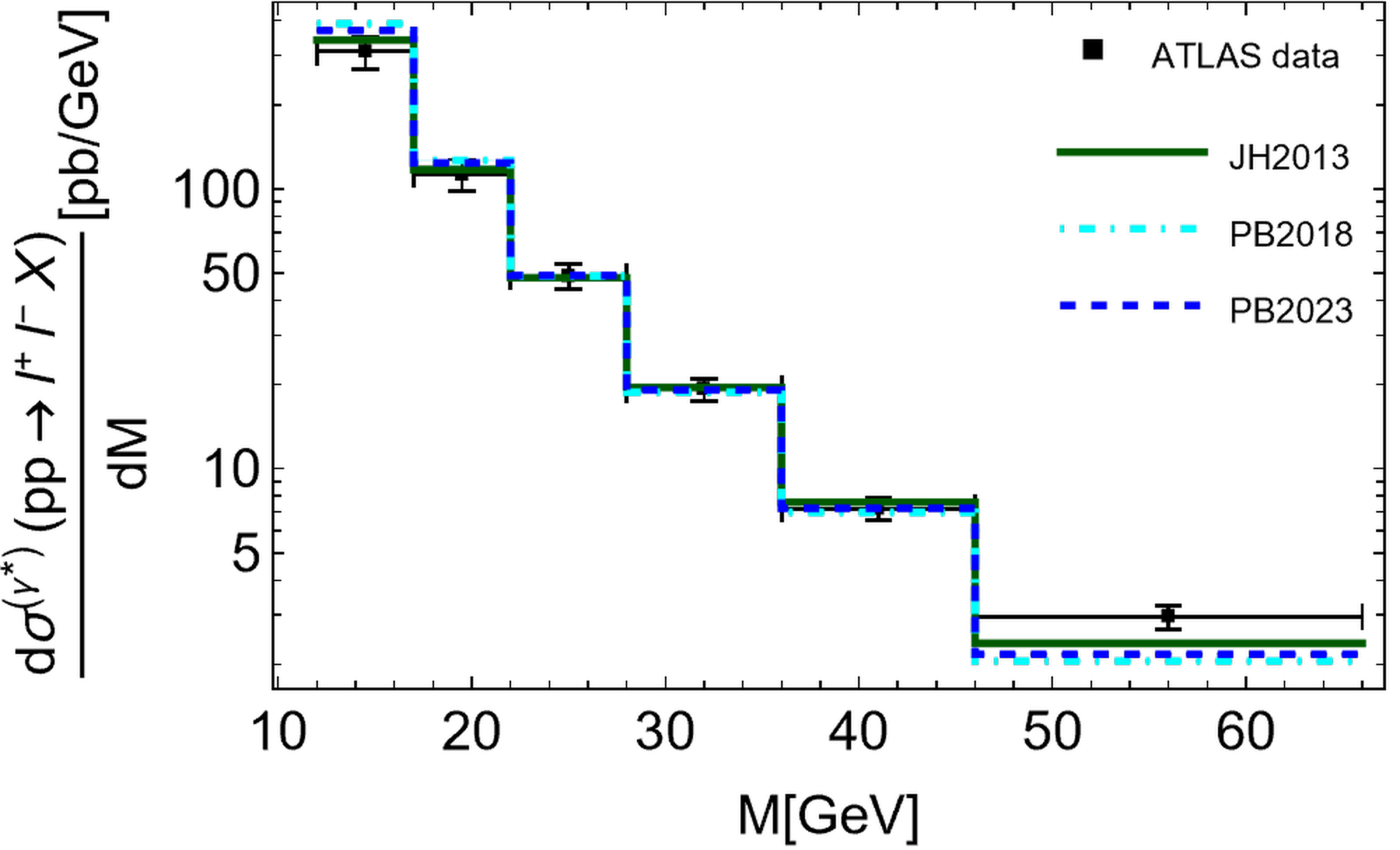}
        \caption{$\dd \sigma^{(\gamma^*)}/\dd M$ for TMD models derived from the CCFM evolution equation.}
        \label{CS CCFM plot}
    \end{subfigure}
    ~
    \begin{subfigure}[h]{0.49\textwidth}
        \centering
        \includegraphics[width=1\textwidth]{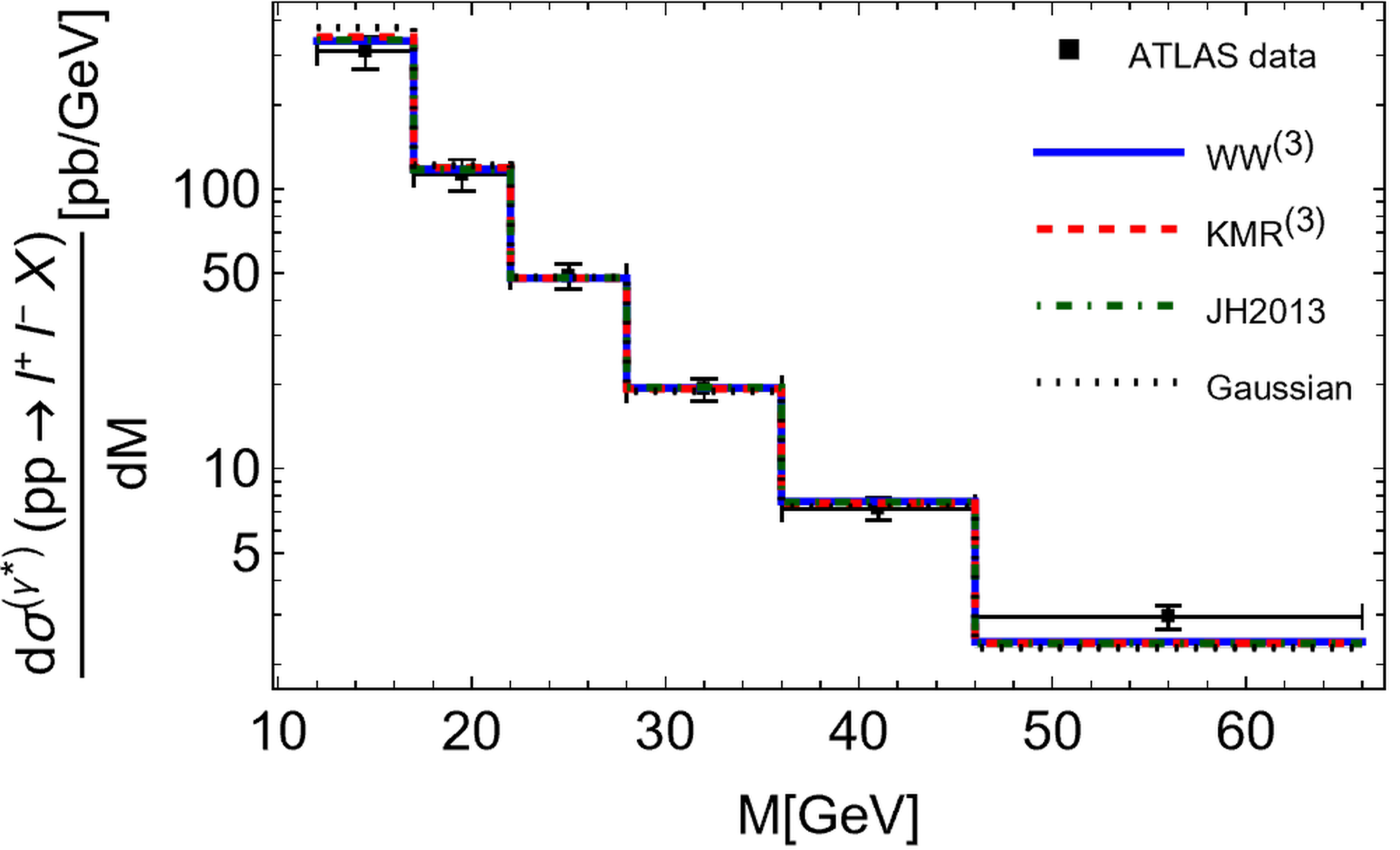}
        \caption{$\dd \sigma^{(\gamma^*)} / \dd M$ for $\mathrm{WW}^{(3)}$, $\mathrm{KMR}^{(3)}$, JH2013 and Gaussian TMD models.}
        \label{CS plot}
    \end{subfigure}
    \caption[]{The Drell--Yan total cross section for $\gamma^*$ exchange in $pp$ collision at the LHC at $\sqrt{S} = \qty{7}{\TeV}$. We compared the ATLAS data \cite{ATLAS:2014ape} to the cross sections calculated with different gluon TMD models: \begin{enumerate*}[label=(\alph*),itemjoin={{, }}]
        \item the KMR-based TMD models
        \item the WW-based TMD models
        \item the CCFM-driven TMD models
        \item $\mathrm{KMR}^{(3)}$, $\mathrm{WW}^{(3)}$, JH2013, and Gaussian models as the models with the best overall data description (including the structure functions analysis) within each model family.
    \end{enumerate*}}
    \label{CS plots}
\end{figure*}

\begin{table}[h!]
\centering
\begin{tabular}{c|cccccccccc}
\hline
\hline
 {\small Model} & {\small $\text{WW}^{(1)}$} & {\small $\text{WW}^{(2)}$} & {\small $\text{WW}^{(3)}$} & {\small $\text{KMR}^{(1)}$} & {\small $\text{KMR}^{(2)}$} & {\small $\text{KMR}^{(3)}$} & {\small JH2013} & {\small PB2018} & {\small PB2023} & {\small Gauss.} \\
\hline
 {\small $N_{\mathrm{TMD}}$} &  2.0848 & 1.3880 & 0.9484 & 2.6021 & 1.7619 & 1.2243 & 0.8728 & 2.2017 & 2.6453 & 0.4158 \\
 \hline
 \hline
\end{tabular}
\caption{Normalization constants for different gluon TMD models.}
\label{N values}
\end{table}

We calculated the cross section $\cfrac{\dd \sigma^{(\gamma^*)}(pp \to l^+ l^- X)}{\dd M}$ averaged over the dilepton mass bins, in the mass regions where the contribution from the $Z^0$ boson is negligible. The cross sections for the normalized TMDs are plotted in Figure \ref{CS plots} and the normalization constants are given in Table \ref{N values}. The reduced $\chi^2$ values, given in Table \ref{CS chi2 values}, for this cross section were calculated according to the formula
\begin{equation}
\label{CS chi2}
    \chi^2 = \frac{1}{N_\mathrm{pt}-1} \sum_{i=1}^{N_\mathrm{pt}} \frac{1}{\sigma_\mathrm{data}^2} \left[\lr{\frac{\dd \sigma^{(\gamma^*)}}{\dd M}}_\mathrm{th} - \lr{\frac{\dd \sigma^{(\gamma^*)}}{\dd M}}_\mathrm{data}\right]^2,
\end{equation}
where $N_\mathrm{pt} = 6$ and we fitted the overall normalization, $\mathrm{th}/\mathrm{data}$ corresponds to the theoretical and experimental values respectively.

\begin{table}[h!]
\centering
\mbox{}\clap{
    \begin{tabular}{c|ccccccccccc}
    \hline
    \hline
    {\small Model} & {\small $\text{WW}^{(1)}$} & {\small $\text{WW}^{(2)}$} & {\small $\text{WW}^{(3)}$} &{\small  WW} & {\small $\text{KMR}^{(1)}$} & {\small $\text{KMR}^{(2)}$} & {\small $\text{KMR}^{(3)}$} & {\small JH2013} & {\small PB2018} & {\small PB2023} & {\small Gauss.} \\
    \hline
    {\small $\chi^2$} & 3.02 & 1.82 & 0.98 & 0.53 & 3.08 & 1.90 & 1.13 & 1.07 & 3.04 & 2.10 & 1.74 \\
    \hline
    \hline
    \end{tabular}
}
\caption{The reduced $\chi^2$ values for different TMD models.}
\label{CS chi2 values}
\end{table}

In the calculations, we applied the $K$-factor, which partially accounts for resummed higher-order QCD corrections. Use of this factor with an approximate form $K = \exp(\pi C_F \alpha_s\lr{\mu_q^2}/2)$ with the optimal choice of the scale $\mu_q = \lr{q_T M^2}^{1/3}$ in the DY cross section calculation was motivated in \cite{Kulesza:1999gm, Watt:2003vf}. In our analysis, we simplified the calculations by assuming an average constant value of $K = 1.5$ instead of the full form depending on $q_T$ and $M$, as was done in Ref.\ \cite{Motyka:2016lta}.

The behavior of the KMR-based and WW'-based  TMDs is very similar. Both show a too steep decrease with increasing $M$, resulting in a slight discrepancy at the ends of the mass spectrum. This problem becomes less significant with stronger $x$ rescaling. A similar discrepancy occurs for the JH models. Table \ref{CS chi2 values} indicates that the WW model comes closest to describing the DY total cross section data, while models with the best structure function data description, i.e.\ $\text{WW}^{(3)}$, $\text{KMR}^{(3)}$ and JH2013, have similar cross section reduced $\chi^2$ values close to 1.

\subsection{The Drell--Yan Structure Functions}
\label{Results DY SF}
\begin{figure*}[h!]
    \begin{subfigure}[ht]{0.49\textwidth}
        \centering
        \includegraphics[width=1\textwidth]{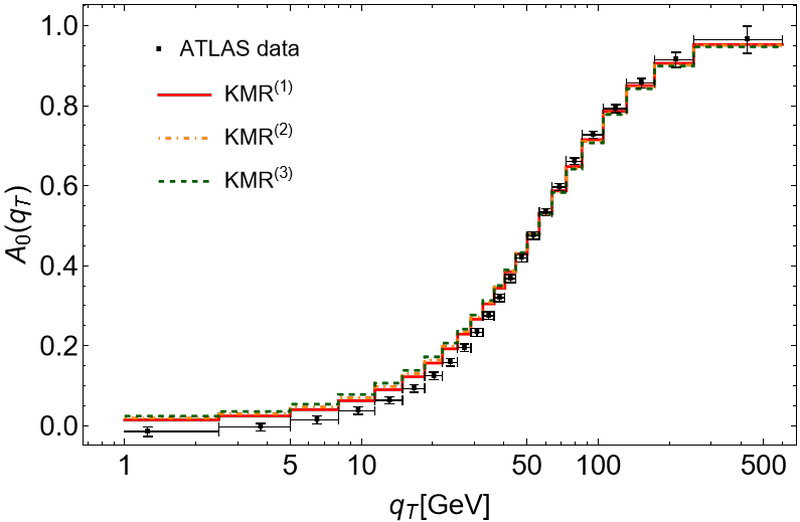}
        \caption{}
        \label{A0 plot KMR}
    \end{subfigure}
    ~
    \begin{subfigure}[h]{0.49\textwidth}
        \centering
        \includegraphics[width=1\textwidth]{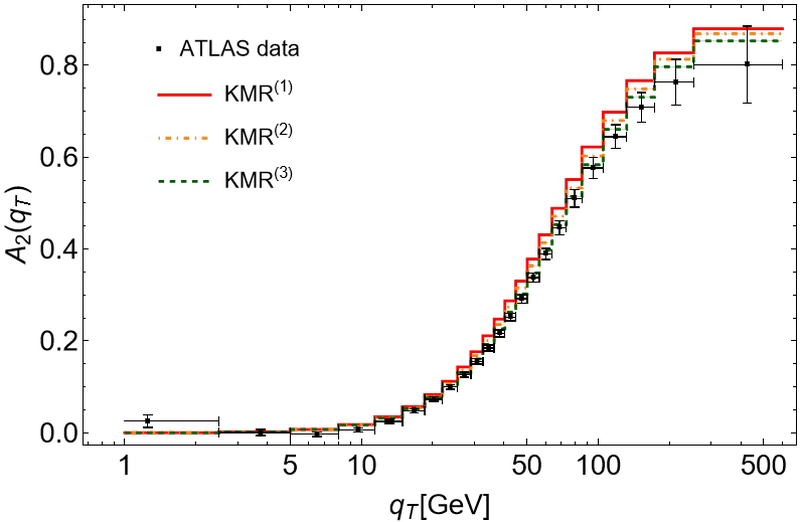}
        \caption{}
        \label{A2 plot KMR}
    \end{subfigure}
    ~
    \begin{subfigure}[h]{0.49\textwidth}
        \centering
        \includegraphics[width=1\textwidth]{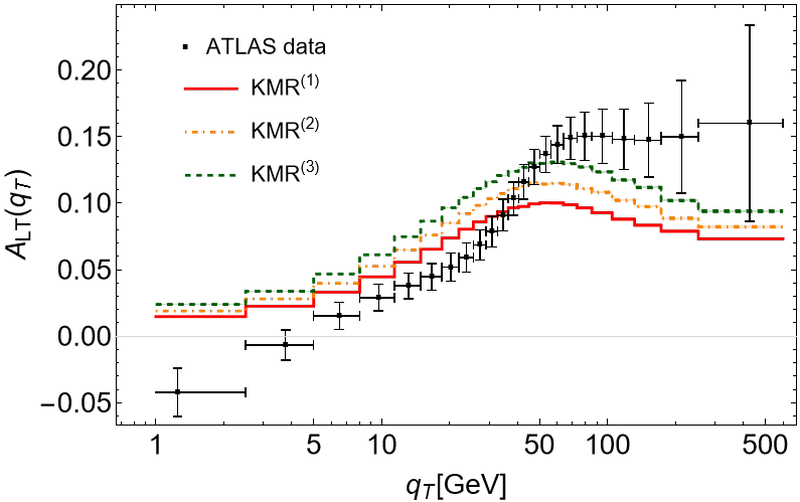}
        \caption{}
        \label{ALT plot KMR}
    \end{subfigure}
    ~
    \begin{subfigure}[h]{0.49\textwidth}
        \centering
        \includegraphics[width=1\textwidth]{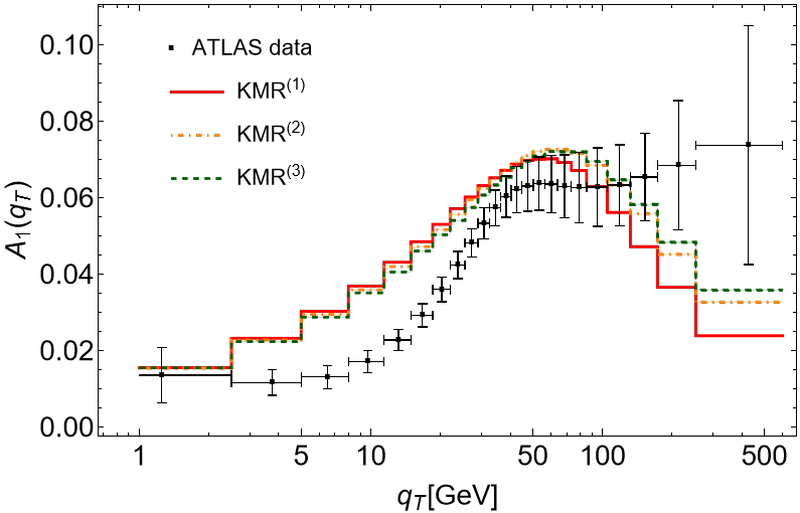}
        \caption{}
        \label{A1 plot KMR}
    \end{subfigure}
    ~
    \begin{subfigure}[h]{0.49\textwidth}
        \centering
        \includegraphics[width=1\textwidth]{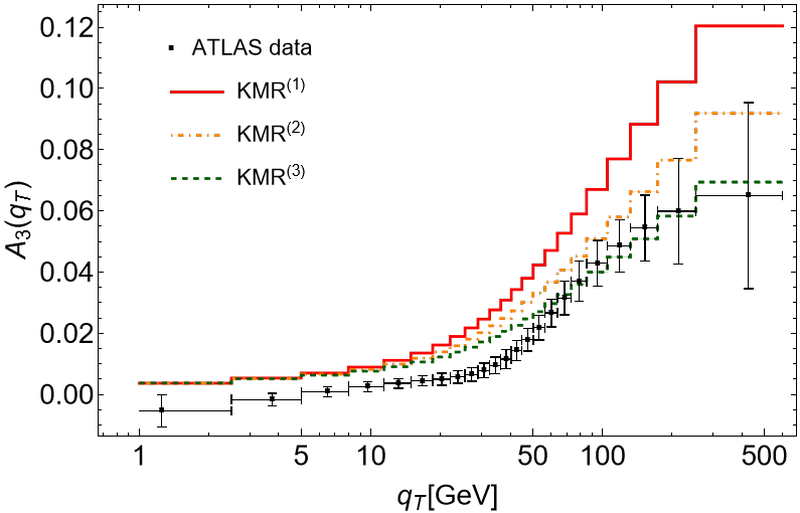}
        \caption{}
        \label{A3 plot KMR}
    \end{subfigure}
    ~
    \begin{subfigure}[h]{0.49\textwidth}
        \centering
        \includegraphics[width=1\textwidth]{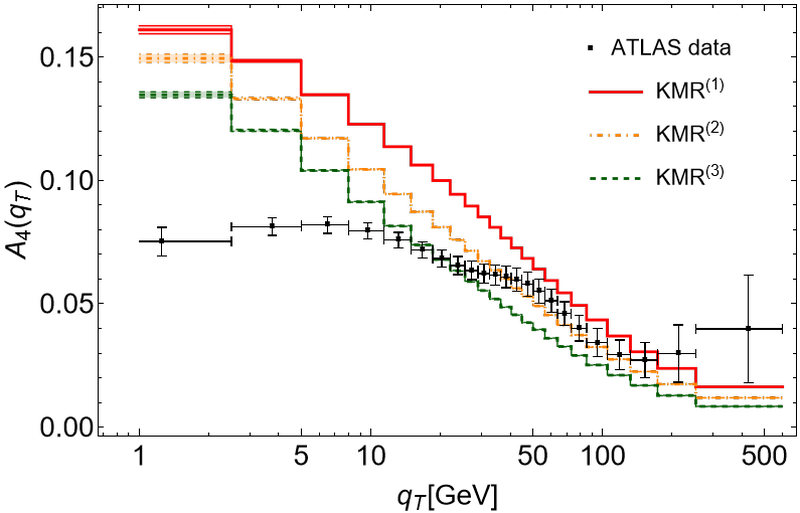}
        \caption{}
        \label{A4 plot KMR}
    \end{subfigure}
    \caption[]{Comparison of the structure functions: \begin{enumerate*}[label=(\alph*),itemjoin={{, }}]
        \item $A_0$
        \item $A_2$
        \item $A_{LT}$
        \item $A_1$
        \item $A_3$
        \item $A_4$
    \end{enumerate*} calculated with different modifications of the KMR TMD model with the ATLAS data \cite{ATLAS:2016rnf}. The bands around the curves represent the estimated Monte Carlo integration uncertainties.}
    \label{SF plots KMR}
\end{figure*}
\begin{figure*}[h!]
    \begin{subfigure}[h]{0.49\textwidth}
        \centering
        \includegraphics[width=1\textwidth]{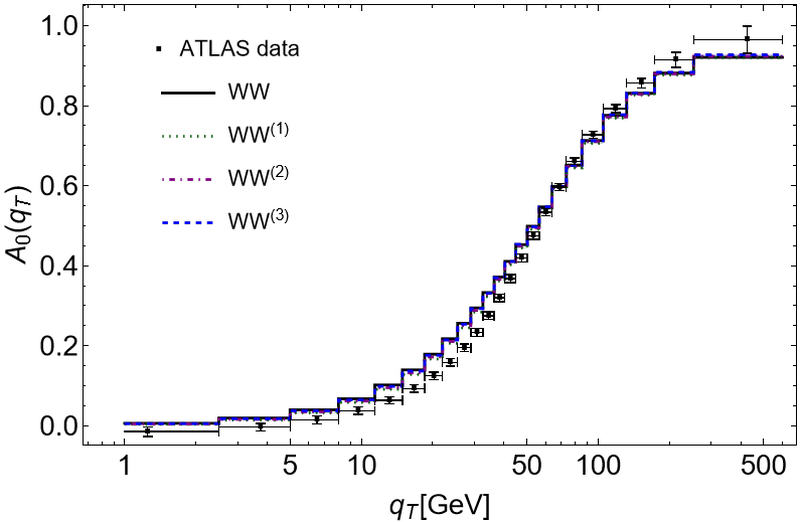}
        \caption{}
        \label{A0 plot WW}
    \end{subfigure}
    ~
    \begin{subfigure}[h]{0.49\textwidth}
        \centering
        \includegraphics[width=1\textwidth]{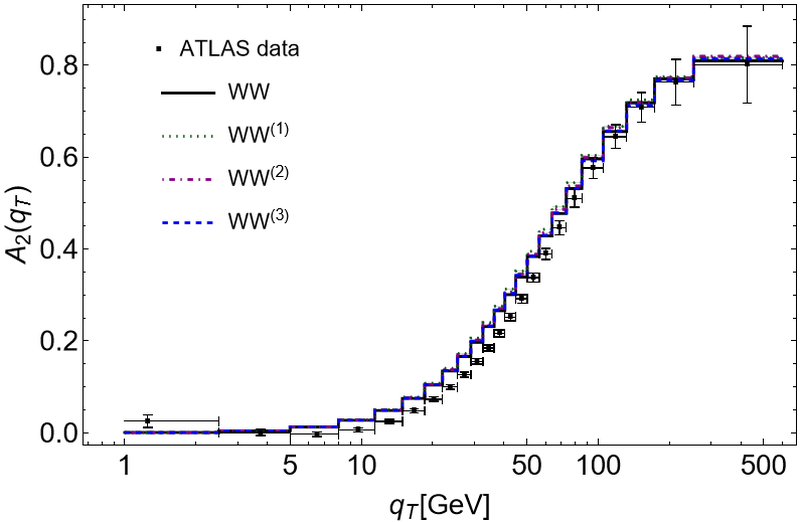}
        \caption{}
        \label{A2 plot WW}
    \end{subfigure}
    ~
    \begin{subfigure}[h]{0.49\textwidth}
        \centering
        \includegraphics[width=1\textwidth]{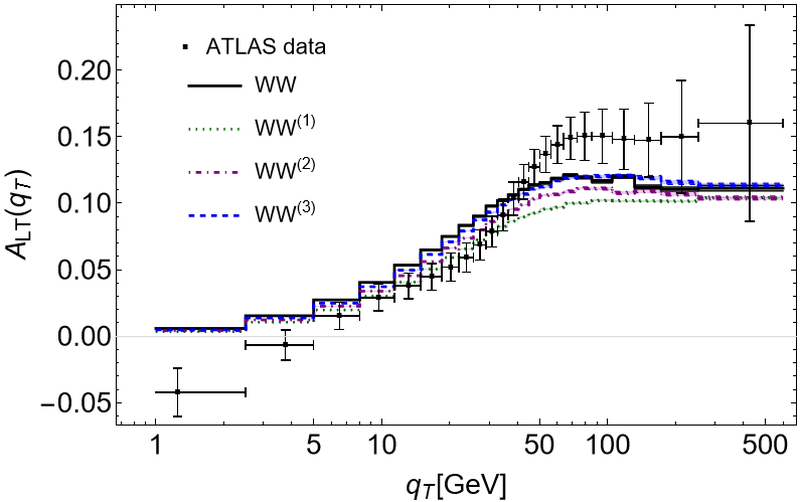}
        \caption{}
        \label{ALT plot WW}
    \end{subfigure}
    ~
    \begin{subfigure}[h]{0.49\textwidth}
        \centering
        \includegraphics[width=1\textwidth]{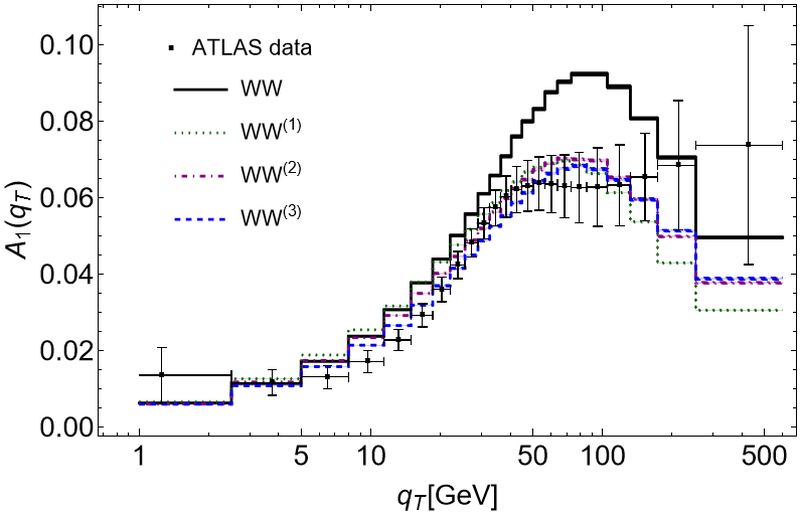}
        \caption{}
        \label{A1 plot WW}
    \end{subfigure}
    ~
    \begin{subfigure}[h]{0.49\textwidth}
        \centering
        \includegraphics[width=1\textwidth]{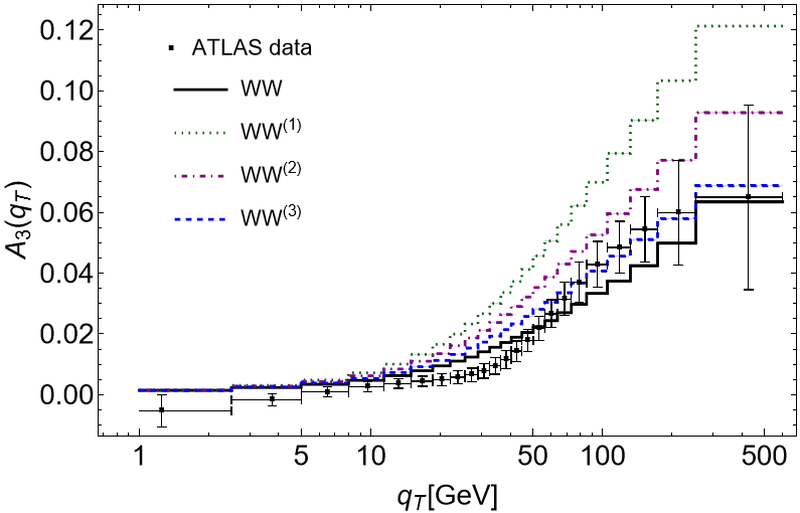}
        \caption{}
        \label{A3 plot WW}
    \end{subfigure}
    ~
    \begin{subfigure}[h]{0.49\textwidth}
        \centering
        \includegraphics[width=1\textwidth]{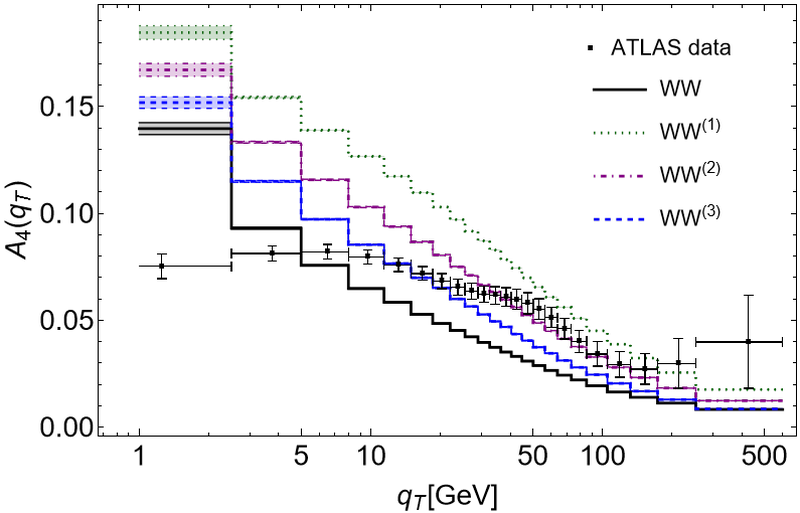}
        \caption{}
        \label{A4 plot WW}
    \end{subfigure}
    \caption[]{Comparison of the structure functions: \begin{enumerate*}[label=(\alph*),itemjoin={{, }}]
        \item $A_0$
        \item $A_2$
        \item $A_{LT}$
        \item $A_1$
        \item $A_3$
        \item $A_4$
    \end{enumerate*} calculated with different modifications of the WW TMD model with the ATLAS data \cite{ATLAS:2016rnf}. The bands around the curves represent the estimated Monte Carlo integration uncertainties.}
    \label{SF plots WW}
\end{figure*}
\begin{figure*}[h!]
    \begin{subfigure}[h]{0.49\textwidth}
        \centering
        \includegraphics[width=1\textwidth]{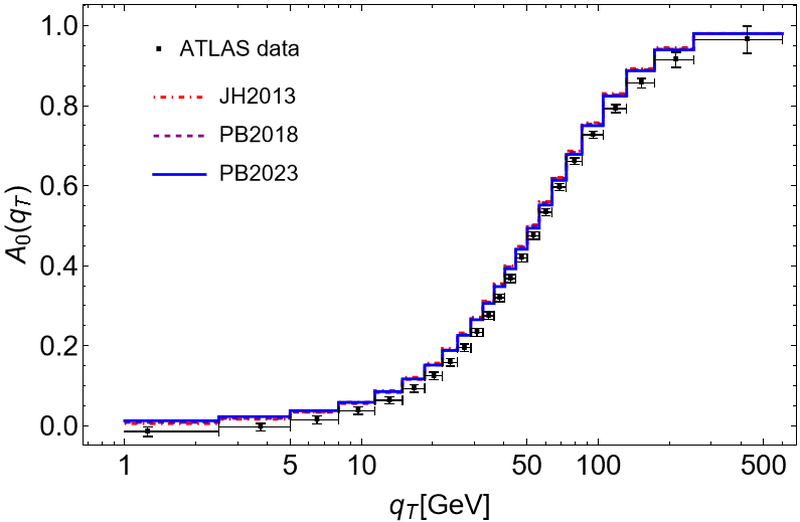}
        \caption{}
        \label{A0 plot CCFM}
    \end{subfigure}
    ~
    \begin{subfigure}[h]{0.49\textwidth}
        \centering
        \includegraphics[width=1\textwidth]{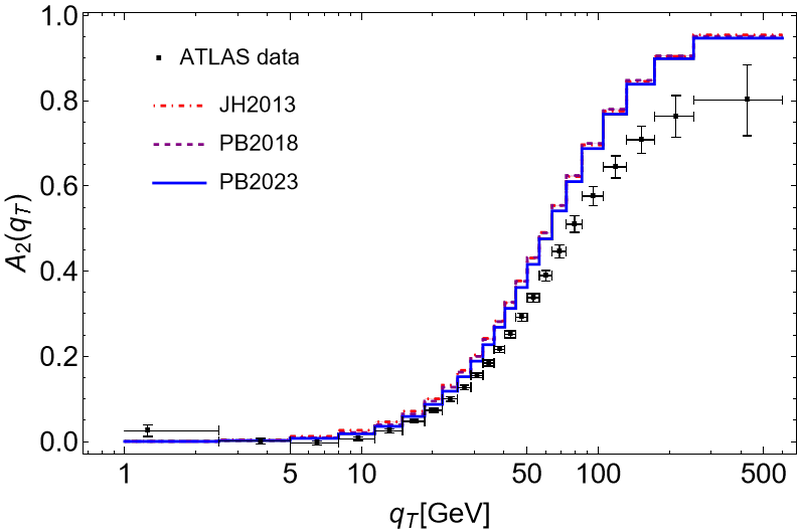}
        \caption{}
        \label{A2 plot CCFM}
    \end{subfigure}
    ~
    \begin{subfigure}[h]{0.49\textwidth}
        \centering
        \includegraphics[width=1\textwidth]{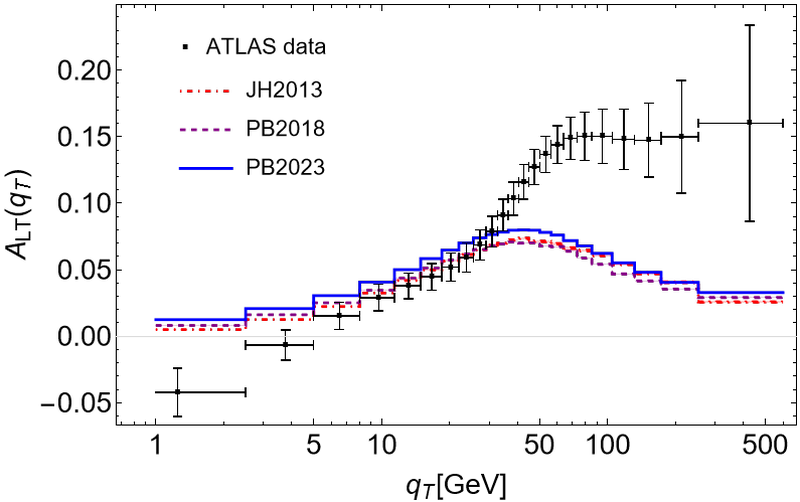}
        \caption{}
        \label{ALT plot CCFM}
    \end{subfigure}
    ~
    \begin{subfigure}[h]{0.49\textwidth}
        \centering
        \includegraphics[width=1\textwidth]{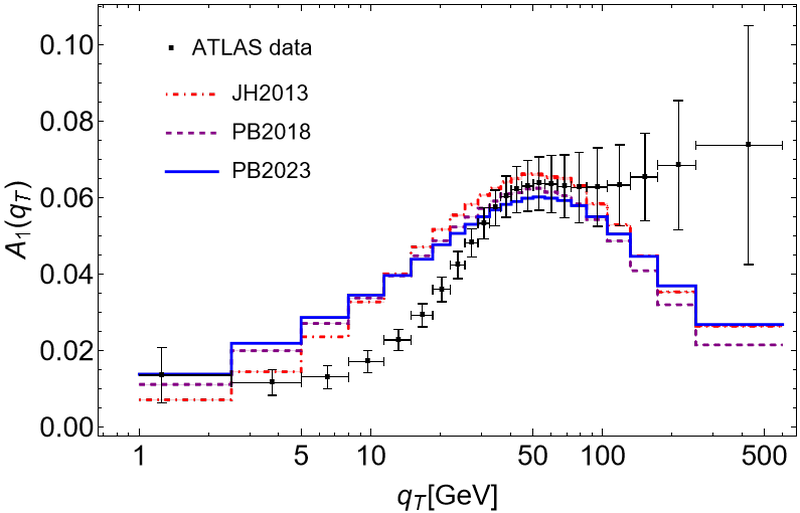}
        \caption{}
        \label{A1 plot CCFM}
    \end{subfigure}
    ~
    \begin{subfigure}[h]{0.49\textwidth}
        \centering
        \includegraphics[width=1\textwidth]{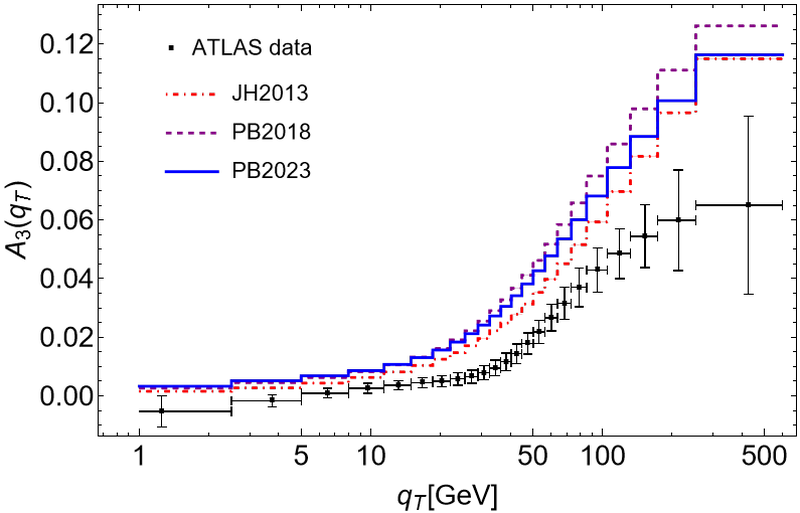}
        \caption{}
        \label{A3 plot CCFM}
    \end{subfigure}
    ~
    \begin{subfigure}[h]{0.49\textwidth}
        \centering
        \includegraphics[width=1\textwidth]{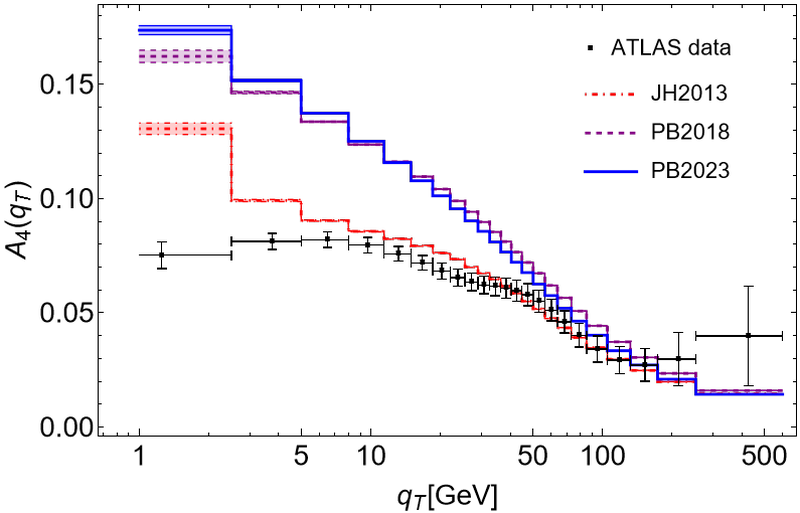}
        \caption{}
        \label{A4 plot CCFM}
    \end{subfigure}
    \caption[]{Comparison of the structure functions: \begin{enumerate*}[label=(\alph*),itemjoin={{, }}]
        \item $A_0$
        \item $A_2$
        \item $A_{LT}$
        \item $A_1$
        \item $A_3$
        \item $A_4$
    \end{enumerate*} calculated with the JH TMD models with the ATLAS data \cite{ATLAS:2016rnf}. The bands around the curves represent the estimated Monte Carlo integration uncertainties.}
    \label{SF plots CCFM}
\end{figure*}
\begin{figure}[h!]
    \begin{subfigure}[h]{0.49\textwidth}
        \centering
        \includegraphics[width=1\linewidth]{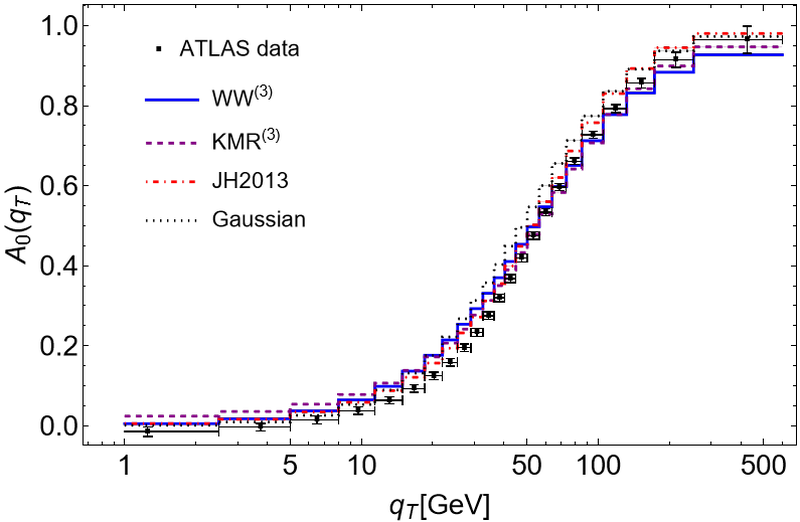}
        \caption{}
        \label{A0 plot}
    \end{subfigure}
    ~
    \begin{subfigure}[h]{0.49\textwidth}
        \centering
        \includegraphics[width=1\linewidth]{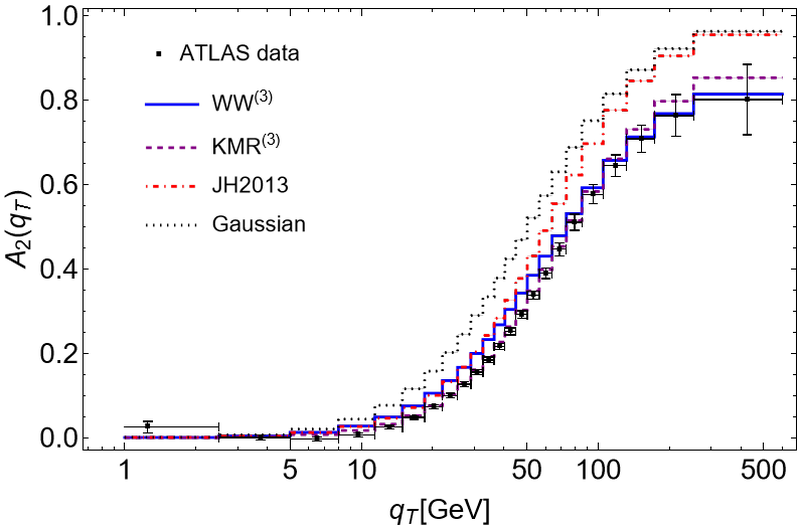}
        \caption{}
        \label{A2 plot}
    \end{subfigure}
    ~
    \begin{subfigure}[h]{0.49\textwidth}
        \centering
        \includegraphics[width=1\linewidth]{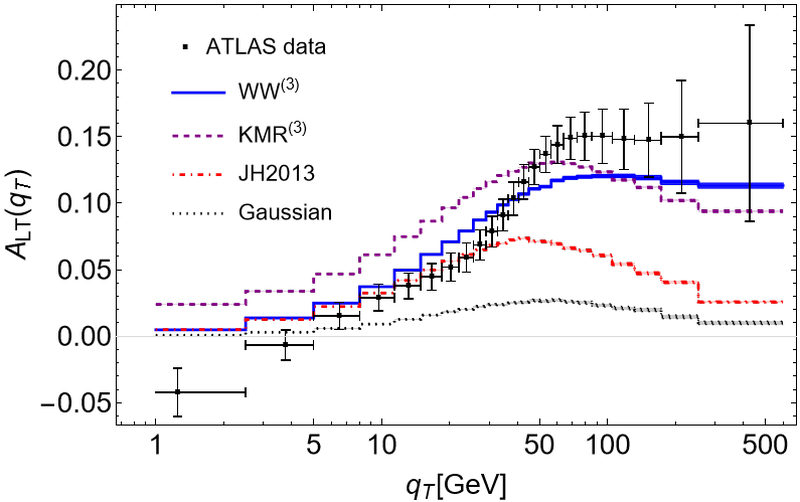}
        \caption{}
        \label{ALT plot}
    \end{subfigure}
    ~
    \begin{subfigure}[h]{0.49\textwidth}
        \centering
        \includegraphics[width=1\linewidth]{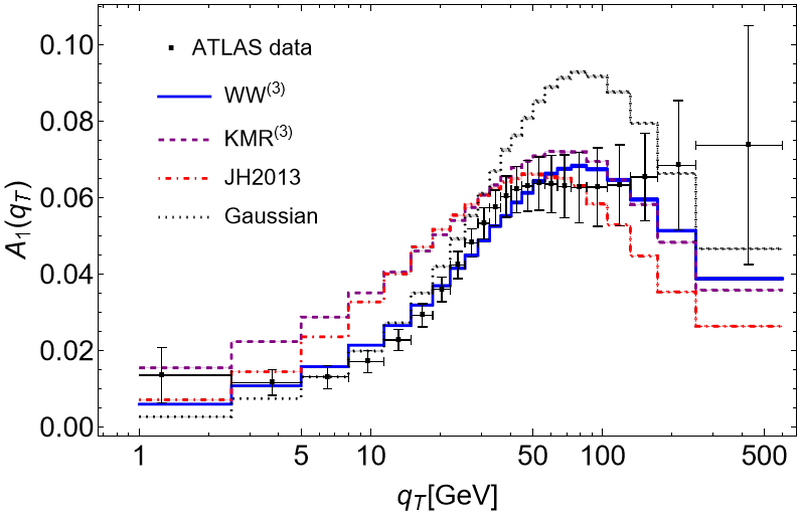}
        \caption{}
        \label{A1 plot}
    \end{subfigure}
    ~
    \begin{subfigure}[h]{0.49\textwidth}
        \centering
        \includegraphics[width=1\linewidth]{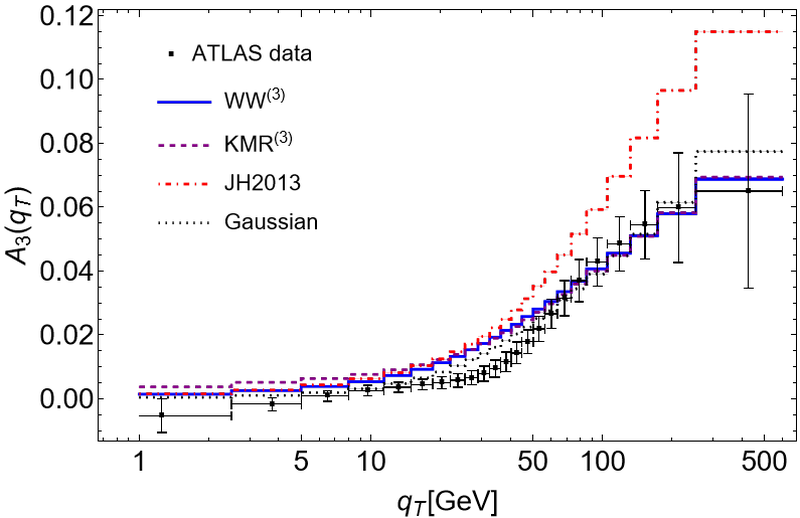}
        \caption{}
        \label{A3 plot}
    \end{subfigure}
    ~
    \begin{subfigure}[h]{0.49\textwidth}
        \centering
        \includegraphics[width=1\linewidth]{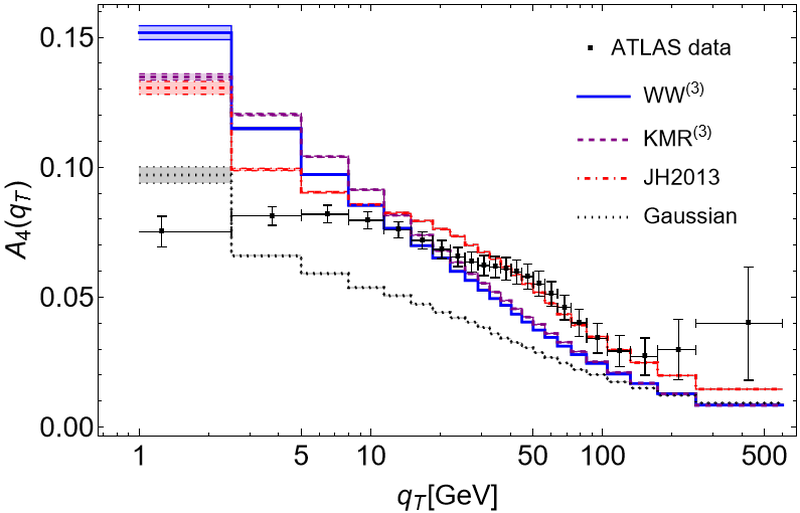}
        \caption{}
        \label{A4 plot}
    \end{subfigure}
    \caption[]{Comparison of the structure functions: \begin{enumerate*}[label=(\alph*),itemjoin={{, }}]
        \item $A_0$
        \item $A_2$
        \item $A_{LT}$
        \item $A_1$
        \item $A_3$
        \item $A_4$
    \end{enumerate*} calculated with different TMD models with ATLAS data. For this purpose, we used $\text{KMR}^{(3)}$, $\text{WW}^{(3)}$, the JH2013 model, and the Gaussian model as representatives of each TMD class that provides the best description of the data within each class. The bands around the curves represent the estimated Monte Carlo integration uncertainties.}
    \label{SF plots}
\end{figure}

In our study, we compared the KMR-based models with one another in Figure \ref{SF plots KMR}, the WW-based models separately in Figure \ref{SF plots WW}, and the JH models in Figure \ref{SF plots CCFM}. Then we compared the representatives with minimal reduced $\chi^2$ in each of the above-mentioned model families with one another and with the Gaussian model. This comparison is shown in Figure \ref{SF plots}.

The $A_0$ function is described very well by all of the models, with minor variations across TMD model families. The most problematic region for all the models is $q_T < \qty{50}{\GeV}$, where we can see a small gap between data and predictions. Although it is barely visible in the plots, this region largely contributes to the reduced $\chi^2$ as the errors are relatively small and points are quite densely distributed there. The $x$ rescaling for WW- and KMR-based models increases this effect.

The $A_2$ function, on the other hand, provides a much clearer distinction between the model families. Here, the WW-based models are closest to the data at large $q_T$, and there is no significant difference between them for this observable. All the other models overestimate the data in that region. The KMR-based family overestimation, however, does not take the values outside the error bars very far. Here we see significant improvement as we rescale the $x$ dependence more. In the region $\qty{5}{\GeV} < q_T < \qty{40}{\GeV}$, one notices a small gap between the data and theory for almost all models except the KMR-based models. This region accounts for the main contribution to the reduced $\chi^2$ for the WW-based and Gaussian models, similarly to the case of $A_0$, because the gap is more significant than for the other models, the errors are relatively small, and the points are quite densely distributed.

The features of the $A_0$ and $A_2$ described above manifest themselves in the Lam--Tung combination $A_{LT}$. In the problematic region of $q_T < \qty{50}{\GeV}$, the $A_{LT}$ is overestimated for all the models except for the Gaussian TMD. This effect is the most visible for the KMR-based models, where the deviation from data is present only for $A_0$, while for the WW-based models, the effects in both functions cancel out. The deviation from data for large $q_T$ is clearly visible for the JH models and especially for the Gaussian model, and it originates from a large discrepancy in $A_2$ at high $q_T$ values.

In the case of function $A_1$, all models show a clear decline at very large $q_T$, contrary to the data trend. The KMR and JH models overestimate the data significantly for small transverse momenta $q_T < \qty{30}{\GeV}$, which constitutes the dominant contribution to the reduced $\chi^2$. On the other hand, they quite accurately describe the position of the local maximum of this function and approximately its value. The WW'-based models generally provide the best description of this observable with significant improvement with stronger $x$ rescaling. A distinct behavior can be noticed for the WW and Gaussian models, namely, they both highly overestimate the $A_1$ local maximum. The difference between WW and WW'-based models indicates a strong effect of the TMD $x$ dependence for this structure function.

The $A_3$ function is also overestimated at large $q_T$ for the JH models and for the WW' and KMR models without $x$ rescaling. However, the region where the deviation from the data is strongest is roughly $\qty{10}{\GeV} < q_T < \qty{70}{\GeV}$. The experimental values slowly increase up to transverse momentum around $q_T \approx \qty{35}{\GeV}$, where the values start to increase more rapidly. All the predictions show, however, a much more gentle and monotonic increase in this region. The $x$ rescaling has a big impact on this observable.

We also notice a large spread of predictions for the $A_4$ function. Similarly to the other parity-violating function --- $A_3$, the $x$ rescaling leads to large changes in the description for both WW and KMR model families. In all the models, we observe a too steep decline, especially for small values of $q_T$, while the data points show a plateau. This feature is particularly evident in the WW model family, where the convexity of the curve appears to be opposite to that observed in the data and predictions of the other models.

To quantify these observations, we also evaluated a global reduced $\chi^2$ for each TMD model. The $\chi^2$ per number of degrees of freedom is computed with the formula
\begin{equation}
\label{chin2 formula}
    \chi_n^2 = \frac{1}{N_\mathrm{pt}} \sum_{i = 1}^{N_\mathrm{pt}} \frac{\lr{A^{\mathrm{th}}_{n, i} - A^{\mathrm{data}}_{n, i}}^2}{\lr{\sigma_{n, i}^{\mathrm{data}}}^2 + \lr{\sigma_{n, i}^{\mathrm{th}}}^2},
\end{equation}
where $N_\mathrm{pt} = 23$ is the number of points, $A^{\mathrm{th}/\mathrm{data}}_{n, i}$ is a theoretical/experimental value of the function $A_n$ at the $i$-th point and $\sigma_{n, i}^{\mathrm{th}/\mathrm{data}}$ are the theoretical/experimental errors. We assumed a theoretical uncertainty of the order $\alpha_s(M_Z) \simeq 0.12$, that is, a relative error of about $12\%$
\begin{equation}
    \sigma_{n, i}^{\mathrm{th}} = 0.12 \cdot A^{\mathrm{th}}_{n, i}.
\end{equation}

The total reduced $\chi^2$ of the results for each of the gluon TMDs is obtained by averaging the contributions of all non-vanishing structure functions
\begin{equation}
    \chi^2 = \frac{1}{5} \sum_{n=0}^4 \chi_n^2.
\end{equation}
In Table \ref{chi2 global ratios} we show the ratios of the $\chi^2$ values for each model to the minimal one, which is found in the $\mathrm{WW}^{(3)}$ model and is equal to $\chi_\mathrm{min}^2 \coloneqq \min\{\chi^2\} = 2.50$.

\begin{table}[h!]
\centering
\mbox{}\clap{
    \begin{tabular}{c|cccccccccccc}
    \hline
    \hline
    {\small Model} & {\small WW} & {\small $\text{WW}^{(1)}$} & {\small $\text{WW}^{(2)}$} & {\small $\text{WW}^{(3)}$} & {\small $\text{KMR}^{(1)}$} & {\small $\text{KMR}^{(2)}$} & {\small $\text{KMR}^{(3)}$} & {\small JH2013} & {\small PB2018} & {\small PB2023} & {\small Gauss.} \\
    \hline
    {\small $\chi^2/\chi_{\mathrm{min}}^2$} & 1.54 & 1.8 & 1.14 & 1.00 & 1.74 & 1.21 & 1.11 & 1.16 & 1.91 & 1.71 & 2.22 \\
    \hline
    \hline
    \end{tabular}
}
\caption{The ratios of reduced $\chi^2$ values for each TMD to the minimal $\chi^2$ value --- $\chi_\mathrm{min}^2$.}
\label{chi2 global ratios}
\end{table}

\begin{table}[h!]
\centering
\mbox{}\clap{
    \begin{tabular}{c|cccccccccccc}
    \hline
    \hline
    {\small Model} & {\small WW} & {\small $\text{WW}^{(1)}$} & {\small $\text{WW}^{(2)}$} & {\small $\text{WW}^{(3)}$} & {\small $\text{KMR}^{(1)}$} & {\small $\text{KMR}^{(2)}$} & {\small $\text{KMR}^{(3)}$} & {\small JH2013} & {\small PB2018} & {\small PB2023} & {\small Gauss.} \\
    \hline
    {\small $\chi_0^2/\chi_{\mathrm{min}}^2$} & 0.94 & 0.60 & 0.72 & 0.84 & 0.94 & 0.67 & 1.26 & 0.52 & 0.48 & 0.57 & 0.82 \\
    {\small $\chi_1^2/\chi_{\mathrm{min}}^2$} & 0.76 & 0.38 & 0.22 & 0.15 & 1.38 & 1.24 & 1.10 & 0.98 & 1.06 & 1.05 & 0.70 \\
    {\small $\chi_2^2/\chi_{\mathrm{min}}^2$} & 1.25 & 1.46 & 1.36 & 1.29 & 0.50 & 0.36 & 0.27 & 1.56 & 1.25 & 0.88 & 4.35 \\
    {\small $\chi_3^2/\chi_{\mathrm{min}}^2$} & 0.79 & 4.47 & 2.48 & 1.34 & 4.34 & 2.66 & 1.78 & 2.40 & 4.72 & 4.21 & 0.55 \\
    {\small $\chi_4^2/\chi_{\mathrm{min}}^2$} & 3.95 & 2.09 & 0.89 & 1.39 & 1.80 & 0.86 & 1.15 & 0.35 & 2.04 & 1.86 & 4.70 \\
     \hline
     \hline
    \end{tabular}
}
\caption{The ratios of $\chi^2$ per degree of freedom values for each structure function --- $\chi_n^2$ for each TMD model to the minimal global reduced $\chi^2$ value --- $\chi_\mathrm{min}^2$.}
\label{chi2_n global ratios}
\end{table}

\begin{table}[h!]
\centering
\mbox{}\clap{
    \begin{tabular}{c|cccccccccccc}
    \hline
    \hline
    {\small Model} & {\small WW} & {\small $\text{WW}^{(1)}$} & {\small $\text{WW}^{(2)}$} & {\small $\text{WW}^{(3)}$} & {\small $\text{KMR}^{(1)}$} & {\small $\text{KMR}^{(2)}$} & {\small $\text{KMR}^{(3)}$} & {\small JH2013} & {\small PB2018} & {\small PB2023} & {\small Gauss.} \\
    \hline
    {\small $\chi_{LT}^2/\chi_{LT, \mathrm{min}}^2$} & 1.18 & 1.58 & 1.17 & 1.00 & 2.39 & 2.32 & 2.80 & 5.14 & 5.83 & 4.73 & 17.67 \\
    \hline
    \hline
    \end{tabular}
}
\caption{The ratios of the Lam--Tung combination $\chi^2$ values for each TMD to the minimal $\chi_{LT}^2$ value.}
\label{chi2_LT ratios}
\end{table}

As anticipated from the plots, the dominant contribution to the discrepancies comes from the parity-violating structure functions, which results in the wide range of $\chi^2$ values listed in Table \ref{chi2_n global ratios}.

We considered the Lam--Tung combination separately, as it is not an independent structure function. In this case, $\chi^2$ is calculated by an analogous formula to \eqref{chin2 formula}. The results are given in Table \ref{chi2_LT ratios}, where the minimum $\chi_{LT}^2$ value is found again for the $\mathrm{WW}^{(3)}$ model, and is equal to $\chi_{LT, \mathrm{min}}^2 \coloneqq \min\{\chi_{LT}^2\} = 1.43$.
\section{Discussion}

\begin{figure}[h!]
    \begin{subfigure}[h]{0.49\textwidth}
        \centering
        \includegraphics[width=1\linewidth]{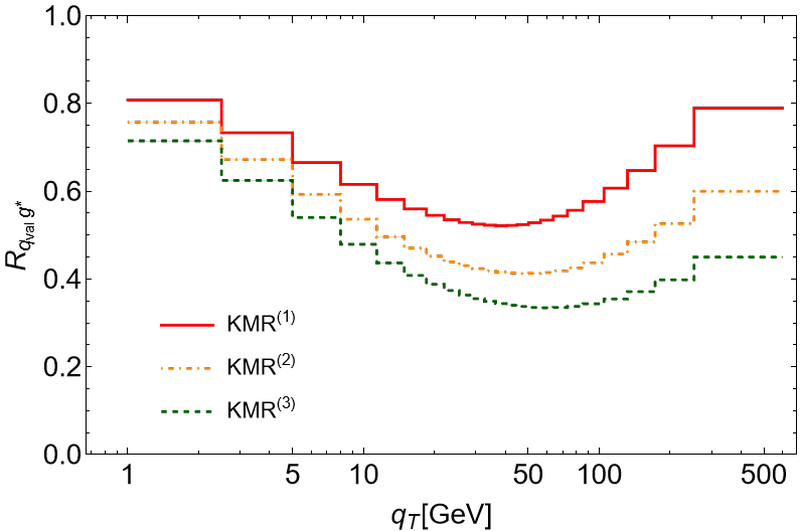}
        \caption{Ratios of the $q_\mathrm{val} g^*$ channel contribution to the total cross section for the KMR-based models.}
        \label{Ratio KMR plot}
    \end{subfigure}
    \hfill
    \begin{subfigure}[h]{0.49\textwidth}
        \centering
        \includegraphics[width=1\linewidth]{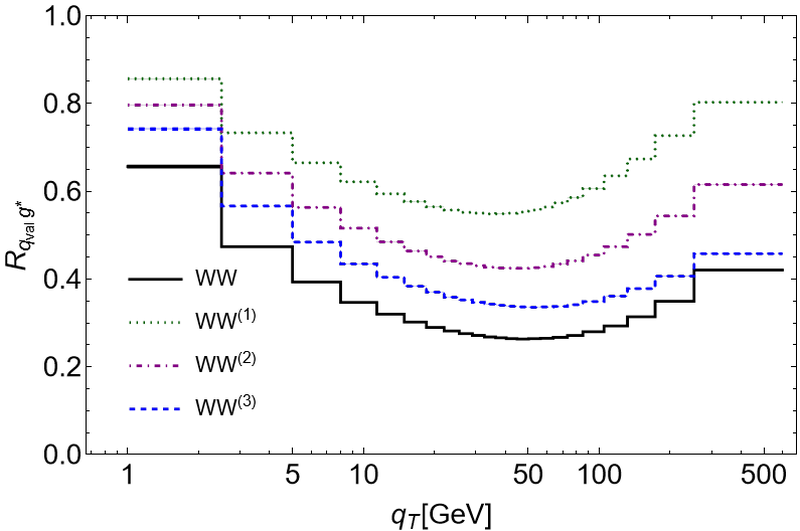}
        \caption{Ratios of the $q_\mathrm{val} g^*$ channel contribution to the total cross section for the WW-based models.}
        \label{Ratio WW plot}
    \end{subfigure}
    \\
    \begin{subfigure}[h]{0.49\textwidth}
        \centering
        \includegraphics[width=1\linewidth]{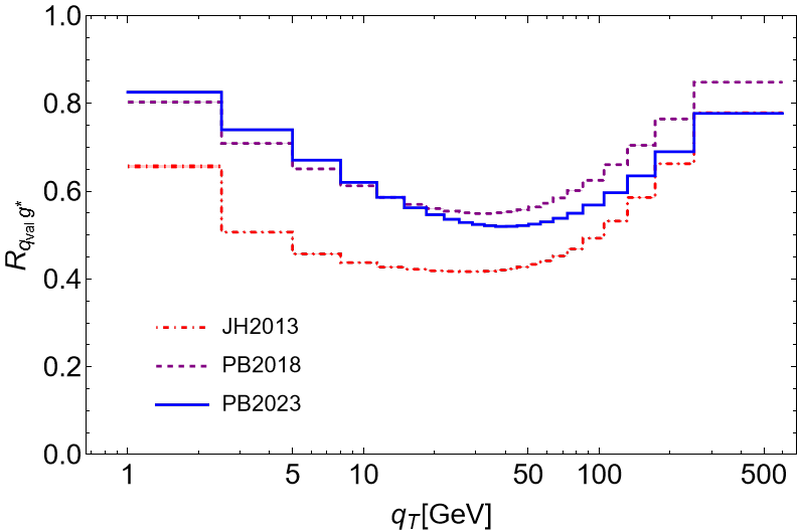}
        \caption{Ratios of the $q_\mathrm{val} g^*$ channel contribution to the total cross section for the CCFM driven models.}
        \label{Ratio CCFM plot}
    \end{subfigure}
    \hfill
    \begin{subfigure}[h]{0.49\textwidth}
        \centering
        \includegraphics[width=1\linewidth]{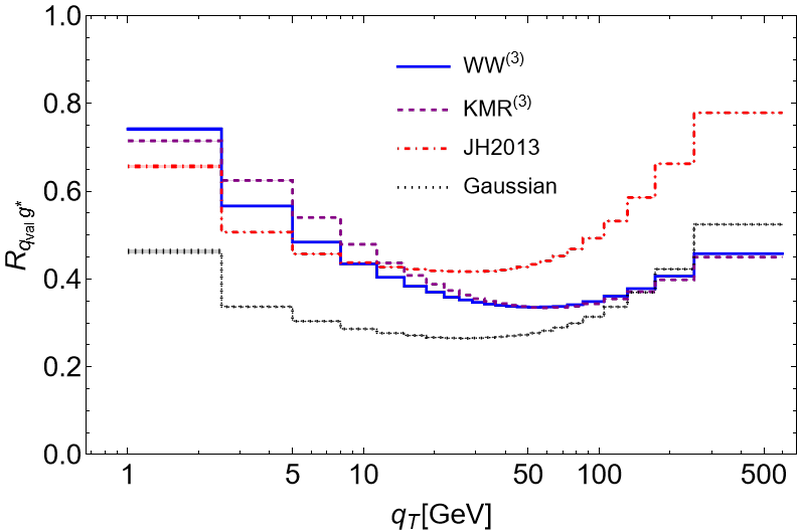}
        \caption{Comparison of ratios of the $q_\mathrm{val} g^*$ channel contribution to the total cross section for the different model families.}
        \label{Ratio plot}
    \end{subfigure}
    \caption{The ratios of the $q_\mathrm{val} g^*$ channel contributions to the total cross section $R_{q_\mathrm{val} g^*}$.}
    \label{Ratio plots}
\end{figure}

After a brief look at the plots in Figure \ref{SF plots}, one recognizes that the structure functions are sensitive to the chosen TMD model over the entire spectrum of $q_T$. Although the main features of each structure function, such as the maximum position and monotonicity, are roughly preserved for each TMD, there is quite a large variation in magnitudes of the theoretical predictions, especially for functions $A_3$ and $A_4$. This variation is mostly due to the difference in the contribution of the $q_\mathrm{val} g^*$ channel to the cross section, as only this channel contributes to these functions. This contribution is quantified by the ratio
\begin{equation}
    \left.
    R_{q_\mathrm{val} g^*} \coloneqq \lr{\frac{\dd \sigma_{(qg^* + g^*q)}}{\dd M \dd q_T}} \right/ \lr{\frac{\dd \sigma}{\dd M \dd q_T}},
\end{equation}
which is plotted for all the TMD models in Figure \ref{Ratio plots}. The different contributions of this channel to the cross section in different models predictions arise from the size of the gluon TMD, which is especially evident when comparing the KMR- and WW-based models. As the gluon $x$ decreases, TMD is enhanced over a wider range of phase space, and consequently, the contribution of the $g^*g^*$ channel to the cross section is larger.

Another effect that occurs because of the change in the contribution of the $q_{\mathrm{val}} g^*$ channel to the total differential cross section is the change of the shape of the function $A_4$ and slightly of the function $A_3$, and the shift of the local maximum of the function $A_1$. The $q_{\mathrm{val}} g^*$ channel contribution is dominant at $q_T = \qty{1}{\GeV}$ for all considered TMD models except of the Gaussian, and decreases noticeably until it reaches the minimum somewhere in the region between $\qty{20}{\GeV}$ to $\qty{70}{\GeV}$ and then slowly grows. If one considers only $\dd \sigma_{(qg^*)}$ in definition \eqref{structure functions definition} for the structure functions $A_3$ and $A_4$, shapes are similar to the shapes visible from the data (apart from the magnitude). However, when one includes the $g^*g^*$ channel, which does not contribute to these functions but whose contribution to the cross section changes significantly with $q_T$, the shape of the discussed functions is altered. Similarly, the shift of the $A_1$ local maximum is related to the contribution of this channel. For the exclusive use of $\dd \sigma_{(qg^*)}$ in \eqref{structure functions definition} the $A_1$ maximum tends to the left whether for the exclusive use of $\dd \sigma_{(g^*g^*)}$ in \eqref{structure functions definition} the $A_1$ maximum tends to the right and the total result is somewhere in between as a weighted average. Therefore, if the $g^*g^*$ channel becomes more significant, the discussed maximum position tends to the higher values of $q_T$.

Table \ref{chi2 global ratios} shows that the model that describes the data best is the $\text{WW}^{(3)}$ model. From Table \ref{chi2_n global ratios} one concludes that the advantage of $\text{WW}^{(3)}$ is due to the rather uniform, relatively good description of all functions and the exceptionally good description of $A_1$. As it can be seen from Table \ref{chi2 global ratios} the models $\text{KMR}^{(3)}$, $\text{WW}^{(2)}$, JH2013 and $\text{KMR}^{(2)}$ give quite close overall results, what suggests that the gluon contribution should be indeed dominant, as these models are the ones with the most dominant $g^* g^*$ channel within each TMD class. The noticeable differences in the description of the structure functions within the WW and KMR families, as well as the differences between the WW'-based models and the WW model, highlight the sensitivity to the $x$ dependence in the $k_T$ factorization approach and its distinction from the collinear approach.

An important note is that the ATLAS data used for comparison with the predictions are much denser in the central part of the $q_T$ spectrum than at its edges. Additionally, the high $q_T$ region is subject to greater uncertainties. Therefore, the middle $q_T$ region is weighted more in the $\chi^2$ test. This feature is described in Section \ref{Results DY SF} and plays a role mostly for the $A_0$, $A_2$, and $A_3$ structure functions.

The behavior of functions $A_3$ and $A_4$ is not fully understood within our approach, and it requires further investigation. One possible solution may be the inclusion of higher-order corrections for the channel $q_{\mathrm{val}} g^*$. In our model, this channel was considered to be subleading due to the dominant contribution of gluons. However, numerical calculations of the contribution of this channel to the full differential cross section clearly show that this assumption is not well fulfilled by some of the TMD models, as can be seen in Figure \ref{Ratio plots}. 

It should also be emphasized that the applied high-energy factorization formalism is well suited for $q_T \sim M$, so it is justified for $q_T \lesssim \qty{100}{\GeV}$, and its use for higher values of $q_T$ can be considered a slight stretching of the formalism. Our description should work better also at higher energies and more central rapidity regions, which gives hope for a good description for future LHC measurements at $\sqrt{S} = \qty{13}{\TeV}$ and higher.
\section{Conclusions}

The DY structure functions are good probes for testing gluon TMD models. In our research, we found clear differences in the descriptions of DY structure functions both between gluon TMD model families and within individual families. The differences in the description are quantified by the reduced $\chi^2$, which can be read from Table~\ref{chi2 global ratios}. They range from 2.5 to 5.56 and are approximately evenly distributed, indicating noticeable differences between the predictions obtained with different TMD models. The description of the individual structure functions, summarized in Table~\ref{chi2_n global ratios}, shows even greater variation between the considered models. For example, we notice an exceptionally good description of the function $A_1$ by the WW'-based models and a comparably good description of the function $A_2$ by the KMR-based models. On the other hand, function $A_3$ is best described by TMD models, for which channel $g^*g^*$ makes the largest contribution in each model family. The situation is slightly different for function $A_4$, but it also distinguishes between individual models within a given model class rather than between the model classes themselves.

The best overall description of the DY structure function is given by the $\text{WW}^{(3)}$ model within our formalism, which also provides the best description of the Lam--Tung combination. This result indicates that the simple WW-like gluon TMD shape in $k_T$ is a good approximation, and can be taken as a leading behavior in $k_T$ in future gluon TMD fits. The differences within the WW-based models and also within the KMR-based models show that the TMD shape in the $x$ variable plays an important role, and that the difference from the collinear PDF $x$ dependence needs to be taken into account in future studies of the gluon TMD model.

Within the CCFM-driven models, the best overall description is given by the oldest model --- JH2013. The difference in the structure function description was additionally reduced by the TMD normalization procedure. The latter models incorporate in the integral kernel a coupling to the full flavor structure that includes sea quarks, which reduces the size of the gluon TMD compared to evolution driven solely by gluons. The procedure used for the JH2013 model might be better suited for the formalism used in our analysis, as in our approach, the sea quarks come exclusively from an initial off-shell gluons, i.e.\ we do not consider processes with initial off-shell (anti)quarks.

Overall, our results demonstrate that the DY structure functions provide insightful probes of a proton transverse momentum internal dynamics. In particular, the results obtained with different gluon TMD models indicate that both the $k_T$ and longitudinal momentum dependence play crucial roles in obtaining results consistent with the experimental measurements of the DY structure functions.

In future work, the above-described formalism can be used to perform a dedicated fit of a gluon TMD to the structure functions data. Another direction for future research could be calculations of NLO corrections and inclusion of additional parton channels with off-shell initial quarks. Such advancements should improve the accuracy of the $k_T$ factorization approach used in this work and are expected to provide further constraints on the gluon TMD. Further investigation of the DY structure functions offers, therefore, a promising direction for advancing our understanding of the internal structure of hadrons.
\section*{Acknowledgements}

We want to thank Leszek Motyka for the introduction to the Drell--Yan process, for many in-depth discussions, and for proofreading the manuscript. We also thank Tomasz Stebel for reading the manuscript and helpful comments. We would also like to thank Krzysztof Golec-Biernat for the discussion on the KMR model. The research has been supported by a grant from the Faculty of Physics, Astronomy and Applied Computer Science under the Strategic Programme Excellence Initiative at Jagiellonian University. This research was supported by the Polish National Science Centre (NCN) grant no. K/NCN/000170, Sonata Bis 12.
\appendix

\section{The Amplitudes} \label{amplitudes qg derivation}

\subsection{The Amplitudes for \texorpdfstring{$q_{\mathrm{val}} g^* \to q V^*$}{q \mathrm{val} g* -> q V*}}

The process studied in this section is $q_{\mathrm{val}}\lr{p_1} g^*\lr{k} \to q\lr{p_2} V^*\lr{q}$. We will consider massive boosted on-shell quarks even though, for all the structure functions calculations, quarks were assumed to be massless. To restore the amplitudes used above, one simply has to take the limits $m_1, m_2 \to 0$ and $\mathbf{p}_{1T} \to \mathbf{0}$. We will apply the techniques and notation presented in \cite{Ferdyan:2024kmx}. The Sudakov decomposition of the initial quark momentum $p_1$ and the gluon momentum $k$ is
\begin{equation}
    p_1 = x_q P_1 + \frac{m_1^2 + \mathbf{p}_{1T}^2}{x_q S} P_2 + p_{1T}, \qquad k = x_g P_2 + k_T.
\end{equation}
where $p_{1T}^2 = -\mathbf{p}_{1T}^2$, then $p_1^2 = m_1^2$ and $k^2 = -\mathbf{k}_T^2$. The outgoing quark momentum is fixed by the momentum conservation as $p_2 = p_1 + k - q$ with the on-shell constraint $p_2^2 = m_2^2$. We are using the high-energy gluon polarization approximation defined in \eqref{nonsense polarization}.

At the tree level, the full contribution to scattering amplitudes comes from two diagrams given by the expressions
\begin{equation}
\begin{split}
    \mathcal{A}_{1, \sigma_1 \sigma_2}^{a, \mu} =& \frac{-i g}{v_1^2 - m_2^2} T^a_{ij} \overline{u}_{\sigma_2}(p_2) \widehat{P}_2 \left( \widehat{v}_1 + m_2 \right) \Gamma_V^\mu u_{\sigma_1}(p_1) =: -i  g T^a_{ij} A_{1, \sigma_1 \sigma_2}^\mu, \\
    \mathcal{A}_{2, \sigma_1 \sigma_2}^{a, \mu} =& \frac{-i g}{v_2^2 - m_1^2} T^a_{ij} \overline{u}_{\sigma_2}(p_2) \Gamma_V^\mu \left( \widehat{v}_2 + m_1 \right) \widehat{P}_2 u_{\sigma_1}(p_1) =: -i g T^a_{ij} A_{2, \sigma_1 \sigma_2}^\mu,
\end{split}
\end{equation}
where $g$ is the strong coupling constant, $T^a$ are generators of the color $SU(N)$ group in the fundamental representation, $u_\sigma(p)$ denotes the Dirac bispinor for fermion with spin $\sigma$ and momentum $p$, we use the notation $\widehat{p} \coloneqq p_\mu \gamma^\mu$ and define the momenta $v_1$, $v_2$ as
\begin{equation*}
    v_1 = p_2 - k, \qquad v_2 = p_1 + k.
\end{equation*}
The scattering amplitude for the process is given by the sum of the above expressions
\begin{equation}
    \mathcal{A}_{\sigma_1 \sigma_2}^{a, \mu} = \mathcal{A}_{1, \sigma_1 \sigma_2}^{a, \mu} + \mathcal{A}_{2, \sigma_1 \sigma_2}^{a, \mu} = -i g T^a_{ij} \lr{A_{1, \sigma_1 \sigma_2}^\mu + A_{2, \sigma_1 \sigma_2}^\mu} =: -i g T^a_{ij} A_{\sigma_1 \sigma_2}^\mu.
\end{equation}
We can further decompose each amplitude into right-handed $R_n^\mu$ and left-handed $L_n^\mu$ parts
\begin{equation}
\label{right-left amplitude decomposition}
    A_{n, \sigma_1 \sigma_2}^\mu = \lr{v^V_q + a^V_q} R_{n, \sigma_1 \sigma_2}^\mu + \lr{v^V_q - a^V_q} L_{n, \sigma_1 \sigma_2}^\mu,
\end{equation}
where in $R_n^\mu$ we replace $\Gamma_V^\mu$ by $\lr{\id + \gamma_5} \gamma^\mu / 2$ and in $L_n^\mu$ we replace $\Gamma_V^\mu$ by $\lr{\id - \gamma_5}  \gamma^\mu / 2$.

Then the amplitude squared for the $qg^*$ channel is given by
\begin{equation}
\label{Amplitude squared qG}
    \mathcal{M}_{(qg^*)}^{\mu \nu} = \frac{1}{2 N \lr{N^2 - 1}} \sum_{\sigma_1, \sigma_2, a, i, j} \mathcal{A}_{\sigma_1 \sigma_2}^{a, \mu} \overline{\mathcal{A}}_{\sigma_1 \sigma_2}^{a, \nu} = \frac{g^2}{4N} \sum_{\sigma_1, \sigma_2} A_{\sigma_1 \sigma_2}^\mu \overline{A}_{\sigma_1 \sigma_2}^{\nu}.
\end{equation}

\subsubsection{Formulas for Diagrams}

The formulas are presented in the spinorial notation
\begin{equation}
    p^{\dot{A} A} \coloneqq p_\mu \lr{\overline{\sigma}^\mu}^{\dot{A} A} = \myvec{p^{\dot{0} 0} & p^{\dot{0} 1} \\
    p^{\dot{1} 0} & p^{\dot{1} 1}} = \myvec{p^+ & p^{\overline{\perp}} \\
    p^{\perp} & p^- },
\end{equation}
where $\overline{\sigma} = \lr{\id, \bm \sigma}$ for $\bm \sigma$ being the vector of Pauli matrices, and we are using double lightcone coordinates defined as
\begin{equation}
    p^\pm \coloneqq p^0 \pm p^3, \qquad p^{\perp} \coloneqq p^1 + i p^2, \qquad p^{\overline{\perp}} \coloneqq p^1 - i p^2.
\end{equation}
Some formulas are written in terms of basis spinors
\begin{equation}
    o_A = \myvec{0 \\ 1}, \quad \iota_A = \myvec{-1 \\ 0}, \quad o^A = \myvec{1 \\ 0}, \quad \iota^A = \myvec{0 \\ 1}.
\end{equation}

We can reduce the number of independent amplitudes by a factor of two using the symmetries
\begin{equation}
R_{n, -+}^\mu = -\overline{L}_{n,+-}^\mu, \quad L_{n, -+}^\mu = -\overline{R}_{n,+-}^\mu, \quad R_{n, ++}^\mu = \overline{L}_{n, --}^\mu, \quad L_{n, ++}^\mu = \overline{R}_{n, --}^\mu.
\end{equation}
Therefore, for each diagram, we write only $R_{n, --}^\mu, L_{n, --}^\mu, R_{n, +-}^\mu$ and $L_{n, +-}^\mu$.

The first diagram:
\begin{equation}
\begin{split}
    R_{1, --}^{\dot A A} =& \frac{2\sqrt{S}}{v_1^2 - m_2^2} \frac{p_2^{\dot 0 0}}{\sqrt{p_1^+ p_2^+}} v_1^{\dot A 0} p_1^{\dot 0 A}, \\
    L_{1, --}^{\dot A A} =& \frac{2\sqrt{S}}{v_1^2 - m_2^2} \frac{m_1 m_2 p_2^{\dot 0 0}}{\sqrt{p_1^+ p_2^+}} \overline{\iota}^{\dot A} \iota^A,
\end{split}
\end{equation}
\begin{equation}
\begin{split}
    R_{1, +-}^{\dot A A} =& -\frac{2\sqrt{S}}{v_1^2 - m_2^2} \frac{m_2 p_2^{\dot 0 0}}{\sqrt{p_1^+ p_2^+}} \overline{\iota}^{\dot A} p_1^{\dot 0 A}, \\
    L_{1, +-}^{\dot A A} =& \frac{2\sqrt{S}}{v_1^2 - m_2^2} \frac{m_1 p_2^{\dot 0 0}}{\sqrt{p_1^+ p_2^+}} \overline{\iota}^{\dot A} v_1^{\dot 0 A}.
\end{split}
\end{equation}
The second diagram:
\begin{equation}
\begin{split}
    R_{2, --}^{\dot A A} =& \frac{2\sqrt{S}}{v_2^2 - m_1^2} \frac{p_1^{\dot 0 0}}{\sqrt{p_1^+ p_2^+}} p_2^{\dot A 0} v_1^{\dot 0 A}, \\
    L_{2, --}^{\dot A A} =& \frac{2\sqrt{S}}{v_2^2 - m_1^2} \frac{m_1 m_2 p_1^{\dot 0 0}}{\sqrt{p_1^+ p_2^+}} \overline{\iota}^{\dot A} \iota^A,
\end{split}
\end{equation}
\begin{equation}
\begin{split}
    R_{2, +-}^{\dot A A} =& -\frac{2\sqrt{S}}{v_2^2 - m_1^2} \frac{m_2 p_1^{\dot 0 0}}{\sqrt{p_1^+ p_2^+}} \overline{\iota}^{\dot A} v_2^{\dot 0 A}, \\
    L_{2, +-}^{\dot A A} =& \frac{2\sqrt{S}}{v_2^2 - m_1^2} \frac{m_1 p_1^{\dot 0 0}}{\sqrt{p_1^+ p_2^+}} \overline{\iota}^{\dot A} p_2^{\dot 0 A}.
\end{split}
\end{equation}

\subsection{The Amplitudes for \texorpdfstring{$g^* g^* \to q \overline{q} V^*$}{g* g* -> q q V*}}

We list here the amplitudes that were used to define $\mathcal{M}_{(g^* g^*)}^{(r_1 r_2)}$ in \eqref{differential cross section GG}
\begin{subequations}
\label{gg amplitudes}
\begin{gather}
    \mathcal{A}_{1, \sigma_1 \sigma_2}^{ab, \mu} = \frac{-i g^2}{\lr{ v_1^2 - m_4^2} \lr{v_2^2 - m_4^2}} \left(T^a T^b\right)_{ij} \overline{u}_{\sigma_3}(p_3) \Gamma_V^\mu \left(\widehat{v}_1 + m_4 \right) \widehat{P}_1 \left(\widehat{v}_2 + m_4 \right) \widehat{P}_2 v_{\sigma_4} (p_4), \\
    \mathcal{A}_{2, \sigma_1 \sigma_2}^{ab, \mu} = \frac{-i g^2}{\left( v_3^2 - m_3^2 \right) \left( v_2^2 - m_4^2 \right)} \left(T^a T^b\right)_{ij} \overline{u}_{\sigma_3}(p_3) \widehat{P}_1 \left(\widehat{v}_3 + m_3 \right) \Gamma_V^\mu \left(\widehat{v}_2 + m_4 \right) \widehat{P}_2 v_{\sigma_4} (p_4), \\
    \mathcal{A}_{3, \sigma_1 \sigma_2}^{ab, \mu} = \frac{-i g^2}{\left( v_3^2 - m_3^2 \right) \left( v_4^2 - m_3^2 \right)} \left(T^a T^b\right)_{ij} \overline{u}_{\sigma_3}(p_3) \widehat{P}_1 \left(\widehat{v}_3 + m_3 \right) \widehat{P}_2 \left(\widehat{v}_4 + m_3 \right) \Gamma_V^\mu v_{\sigma_4} (p_4), \\
    \mathcal{A}_{4, \sigma_1 \sigma_2}^{ab, \mu} = \frac{-i  g^2}{\left( v_1^2 - m_4^2 \right) \left( v_5^2 - m_4^2 \right)} \left(T^b T^a\right)_{ij} \overline{u}_{\sigma_3}(p_3) \Gamma_V^\mu \left(\widehat{v}_1 + m_4 \right) \widehat{P}_2 \left(\widehat{v}_5 + m_4 \right) \widehat{P}_1 v_{\sigma_4} (p_4), \\
    \mathcal{A}_{5, \sigma_1 \sigma_2}^{ab, \mu} = \frac{-i  g^2}{\left( v_6^2 - m_3^2 \right) \left( v_5^2 - m_4^2 \right)} \left(T^b T^a\right)_{ij} \overline{u}_{\sigma_3}(p_3) \widehat{P}_2 \left(\widehat{v}_6 + m_3 \right) \Gamma_V^\mu \left(\widehat{v}_5 + m_4 \right) \widehat{P}_1 v_{\sigma_4} (p_4), \\
    \mathcal{A}_{6, \sigma_1 \sigma_2}^{ab, \mu} = \frac{-i  g^2}{\left( v_4^2 - m_3^2 \right) \left( v_6^2 - m_3^2 \right)} \left(T^b T^a\right)_{ij} \overline{u}_{\sigma_3}(p_3) \widehat{P}_2 \left(\widehat{v}_6 + m_3 \right) \widehat{P}_1 \left(\widehat{v}_4 + m_3 \right) \Gamma_V^\mu v_{\sigma_4} (p_4), \\
    \mathcal{A}_{7, \sigma_1 \sigma_2}^{ab, \mu} = \frac{ g^2 }{\left( k_1 + k_2 \right)^2 \left( v_1^2 - m_4^2 \right)} f^{abc} T^c_{ij} \overline{u}_{\sigma_3}(p_3) \Gamma_V^\mu \left(\widehat{v}_1 + m_4 \right) \widehat{V}_{\mathrm{eff}} v_{\sigma_4} (p_4), \\
    \mathcal{A}_{8, \sigma_1 \sigma_2}^{ab, \mu} = \frac{ g^2 }{\left( k_1 + k_2 \right)^2 \left( v_4^2 - m_3^2 \right)} f^{abc} T^c_{ij} \overline{u}_{\sigma_3}(p_3) \widehat{V}_{\mathrm{eff}} \left(\widehat{v}_4 + m_3 \right) \Gamma_V^\mu v_{\sigma_4} (p_4),
\end{gather}
\end{subequations}
where
\begin{equation}
\begin{split}
    v_1 =& p_3 + q, \quad v_2 = k_2 - p_4, \quad v_3 = p_3 - k_1, \\
    v_4 =& -p_4 - q, \quad v_5 = k_1 - p_4, \quad v_6 = p_3 - k_2,
\end{split}
\end{equation}
and the form of $V_{\mathrm{eff}}^\mu$ results from the contraction of the Lipatov vertex \cite{Lipatov:1995pn} with the approximated gluon polarizations:
\begin{equation}
\begin{split}
\label{eq:Lipatov_vertex}
    V_{\mathrm{eff}}^\mu =& \frac{S}{2} (k_2 - k_1)^\mu + \left( 2 P_2 \cdot k_1 + \frac{P_1 \cdot P_2}{P_1 \cdot k_2} k_1^2 \right) P_1^\mu - \left( 2 P_1 \cdot k_2 + \frac{P_1 \cdot P_2}{P_2 \cdot k_1} k_2^2 \right) P_2^\mu = \\
    =& \frac{S}{2} (k_2 - k_1)^\mu + \left( x_1 S + \frac{k_1^2}{x_2} \right) P_1^\mu - \left( x_2 S + \frac{k_2^2}{x_1} \right) P_2^\mu.
\end{split}
\end{equation}
These amplitudes were calculated from the spinor-helicity formalism in Ref.\ \cite{Ferdyan:2024kmx}.
The amplitude squared is given by
\begin{equation}
\label{Amplitude squared gg}
    \mathcal{M}_{(g^*g^*)}^{\mu \nu} = \frac{1}{\lr{N^2 - 1}^2} \sum_{\sigma_1, \sigma_2, a, b, i, j} \mathcal{A}_{\sigma_1 \sigma_2}^{ab, \mu} \overline{\mathcal{A}}_{\sigma_1 \sigma_2}^{ab, \nu},
\end{equation}
where
\begin{equation}
    \mathcal{A}_{\sigma_1 \sigma_2}^{ab, \mu} = \sum_{n=1}^8 \mathcal{A}_{n, \sigma_1 \sigma_2}^{ab, \mu}.
\end{equation}

\section{The Electroweak Couplings} \label{EW couplings}

The electroweak couplings for fermions with the $Z^0$ boson can be expressed through the Weinberg mixing angle $\theta_W$. The formulas for leptons ($l$) and quarks ($q$) are
\begin{subequations}
\label{EW couplings formulas}
\begin{gather}
    v^{Z}_{l} = \frac{e}{\sin(2\theta_W)} \lr{-1 + 4 \sin^2{\theta_W}}, \qquad a^Z_l = \frac{e}{\sin(2\theta_W)}, \\
    v^Z_q = \frac{e}{\sin{\theta_W} \cos{\theta_W}} \lr{T_{3q} - 2 e_q \sin^2{\theta_W}}, \qquad a^Z_q = \frac{e}{\sin{\theta_W} \cos{\theta_W}} T_{3q},
\end{gather}
\end{subequations}
where $e$ is the value of electron charge, $e_q$ are charge fractions of quarks ($e_u = 2/3$, $e_d = -1/3$) and $T_3$ is the weak isospin with the component values $T_{3u} = 1/2$ and $T_{3d} = -1/2$. The most recent value of the mixing angle published by Particle Data Group gives a value $\sin^2{\theta_W} = 0.23122(6)$ \cite{ParticleDataGroup:2024cfk}.

\begin{figure}[ht]
        \centering
        \includegraphics[height=6cm]{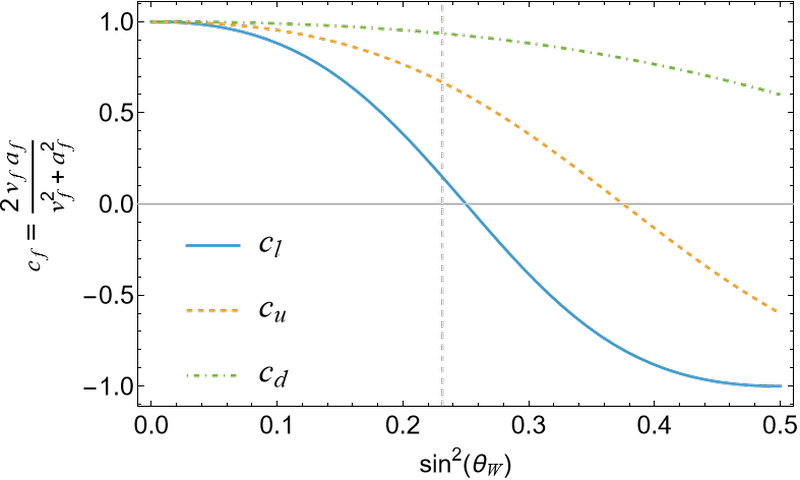}
        \caption{The ratio of electroweak couplings $c_f^Z$ for leptons and quarks. The dashed vertical line corresponds to the measured value of $\sin^2{\theta_W}$.}
        \label{EW couplings plot}
\end{figure}
The accuracy of the mixing angle value is important, especially for parity-breaking structure functions due to the sensitivity of the coupling ratio $c^Z_l$ in the region close to the measured values of $\sin^2{\theta_W}$. There is no such sensitivity in the case of quark couplings, as can be seen in the plot in Fig.\ \ref{EW couplings plot}, where the slope of $c_l$ at the experimental value is quite large compared to the slopes of $c_q$.

\bibliographystyle{unsrt}
\bibliography{refs}

\end{document}